\input harvmac
\input epsf

\newcount\figno
\figno=0
\def\fig#1#2#3{
\par\begingroup\parindent=0pt\leftskip=1cm\rightskip=1cm\parindent=0pt
\baselineskip=12pt
\global\advance\figno by 1
\midinsert
\epsfxsize=#3
\centerline{\epsfbox{#2}}
\vskip 14pt

{\bf Fig. \the\figno:} #1\par
\endinsert\endgroup\par
}
\def\figlabel#1{\xdef#1{\the\figno}}
\def\encadremath#1{\vbox{\hrule\hbox{\vrule\kern8pt\vbox{\kern8pt
\hbox{$\displaystyle #1$}\kern8pt}
\kern8pt\vrule}\hrule}}

\overfullrule=0pt

\noblackbox
\parskip=1.5mm

\def\Title#1#2{\rightline{#1}\ifx\answ\bigans\nopagenumbers\pageno0
\else\pageno1\vskip.5in\fi \centerline{\titlefont #2}\vskip .3in}

\font\caps=cmcsc10

\noblackbox
\parskip=1.5mm



           \def\CO{{\cal O}}


\def\dj{\hbox{d\kern-0.347em \vrule width 0.3em height 1.252ex depth
-1.21ex \kern 0.051em}}

\def\half{{1\over 2}\,}

\def\Tr{{\rm Tr\,}}

\def\pt{\partial}

\def\Dirac{\,\raise.15ex\hbox{/}\mkern-13.5mu D}
\def\dirac{\,\raise.15ex\hbox{/}\kern-.57em \partial}
\def\aslash{\,\raise.15ex\hbox{/}\mkern-13.5mu A}

\def\shalf{{\ifinner {\textstyle {1 \over 2}}\else {1 \over 2} \fi}}
\def\sshalf{{\ifinner {\scriptstyle {1 \over 2}}\else {1 \over 2} \fi}}
\def\sfourth{{\ifinner {\textstyle {1 \over 4}}\else {1 \over 4} \fi}}
\def\sthreehalfs{{\ifinner {\textstyle {3 \over 2}}\else {3 \over 2} \fi}}
\def\sdhalfs{{\ifinner {\textstyle {d \over 2}}\else {d \over 2} \fi}}
\def\sdmtwohalfs{{\ifinner {\textstyle {d-2 \over 2}}\else {d-2 \over 2} \fi}}
\def\sdmasonehalfs{{\ifinner {\textstyle {d+1 \over 2}}\else {d+1 \over 2} \fi}}
\def\sdmasthreehalfs{{\ifinner {\textstyle {d+3 \over 2}}\else {d+3 \over 2} \fi}}
\def\sdmastwohalfs{{\ifinner {\textstyle {d+2 \over 2}}\else {d+2 \over 2} \fi}}


 \lref\adscft{
  J.~M.~Maldacena,
  ``The large N limit of superconformal field theories and supergravity,''
  Adv.\ Theor.\ Math.\ Phys.\  {\bf 2}, 231 (1998)
  [Int.\ J.\ Theor.\ Phys.\  {\bf 38}, 1113 (1999)]
  [arXiv:hep-th/9711200].
 S.~S.~Gubser, I.~R.~Klebanov and A.~M.~Polyakov,
  ``Gauge theory correlators from non-critical string theory,''
  Phys.\ Lett.\  B {\bf 428}, 105 (1998)
  [arXiv:hep-th/9802109].
 E.~Witten,
  ``Anti-de Sitter space and holography,''
  Adv.\ Theor.\ Math.\ Phys.\  {\bf 2}, 253 (1998)
  [arXiv:hep-th/9802150].
  }

\lref\cdl{
 S.~R.~Coleman and F.~De Luccia,
  ``Gravitational Effects On And Of Vacuum Decay,''
  Phys.\ Rev.\  D {\bf 21}, 3305 (1980).
}

\lref\banks{
T.~Banks,
  ``Heretics of the false vacuum: Gravitational effects on and of vacuum decay.
  II,''
  arXiv:hep-th/0211160.
T.~Banks,
  ``Landskepticism or why effective potentials don't count string models,''
  arXiv:hep-th/0412129.}
\lref\bt{
  P.~Breitenlohner and D.~Z.~Freedman,
  ``Stability In Gauged Extended Supergravity,''
  Annals Phys.\  {\bf 144}, 249 (1982).

  P.~Breitenlohner and D.~Z.~Freedman,
  ``Positive Energy In Anti-De Sitter Backgrounds And Gauged Extended
  Supergravity,''
  Phys.\ Lett.\  B {\bf 115}, 197 (1982).
}

\lref\abde{
L.~F.~Abbott and S.~Deser,
  ``Stability Of Gravity With A Cosmological Constant,''
  Nucl.\ Phys.\  B {\bf 195}, 76 (1982).
  }
  
\lref\rgf{
E.~Farhi and A.~H.~Guth,
  ``An Obstacle to Creating a Universe in the Laboratory,"   Phys.\ Lett.\  B {\bf 183}, 149 (1987). \
E.~Farhi, A.~H.~Guth and J.~Guven,
  ``Is it Possible to Create a Universe in the Laboratory by Quantum Tunneling?." 
  		  Nucl.\ Phys.\  B {\bf 339}, 417 (1990).
}

\lref\guthblau{
S.~K.~Blau, E.~I.~Guendelman and A.~H.~Guth,
  ``The Dynamics of False Vacuum Bubbles,''
  Phys.\ Rev.\  D {\bf 35}, 1747 (1987).
}

\lref\tkachev{
 V.~A.~Berezin, V.~A.~Kuzmin and I.~I.~Tkachev,
  ``Thin Wall Vacuum Domains Evolution,''
  Phys.\ Lett.\  B {\bf 120}, 91 (1983).
  V.~A.~Berezin, V.~A.~Kuzmin and I.~I.~Tkachev,
  ``Dynamics of Bubbles in General Relativity,''
  Phys.\ Rev.\  D {\bf 36}, 2919 (1987).}
  
  \lref\aurilia{
  A.~Aurilia, G.~Denardo, F.~Legovini and E.~Spallucci,
  ``Vacuum Tension Effects On The Evolution Of Domain Walls In The Early
  Universe,''
  Nucl.\ Phys.\  B {\bf 252}, 523 (1985).}
  
  \lref\tkachevq{
  V.~A.~Berezin, V.~A.~Kuzmin and I.~I.~Tkachev,
  ``O(3) invariant processes at false vacuum decay in general relativity," 
  Int.\ J.\ Mod.\ Phys.\  A {\bf 5}, 4639 (1990).}

\lref\shenki{
B.~Freivogel, V.~E.~Hubeny, A.~Maloney, R.~C.~Myers, M.~Rangamani and S.~Shenker,
  ``Inflation in AdS/CFT,''
  JHEP {\bf 0603}, 007 (2006)
  [arXiv:hep-th/0510046].
}

\lref\israel{
W.~Israel,
  ``Singular hypersurfaces and thin shells in general relativity,''
  Nuovo Cim.\  B {\bf 44S10}, 1 (1966)
  [Erratum-ibid.\  B {\bf 48}, 463 (1967\ NUCIA,B44,1.1966)].
}

\lref\javi{J.LF. Barb\'on and J. Mart\'{\i}nez-Mag\'an, in preparation.}

\lref\susy{
S.~Weinberg,
  ``Does Gravitation Resolve The Ambiguity Among Supersymmetry Vacua?,''
  Phys.\ Rev.\ Lett.\  {\bf 48}, 1776 (1982).
}

\lref\rey{
M. Cvetic, S. Griffies , S-J. Rey, 
 ``Static domain walls in N=1 supergravity",
Nucl. Phys. B{\bf 381}, 301-328 (1992),  
 [arXiv:hep-th/9201007]. 
M. Cvetic, S. Griffies, Soo-Jong Rey,
``Nonperturbative stability of supergravity and superstring vacua", 
 Nucl \ Phys. \ B{\bf 389}, 3-24 (1993)
[arXiv:hep-th/9206004]. 
M. Cvetic, F. Quevedo, S-J. Rey,
``Stringy domain walls and target space modular invariance." 
Phys. \ Rev.\ Lett.\ {\bf 67},1836-1839 (1991).
}

\lref\susyr{A.~Ceresole, G.~Dall'Agata, A.~Giryavets, R.~Kallosh and A.~D.~Linde,
  ``Domain walls, near-BPS bubbles, and probabilities in the landscape,''
  Phys.\ Rev.\  D {\bf 74}, 086010 (2006)
  [arXiv:hep-th/0605266].
  M.~Dine, G.~Festuccia and A.~Morisse,
  ``The Fate of Nearly Supersymmetric Vacua,''
  JHEP {\bf 0909}, 013 (2009)
  [arXiv:0901.1169 [hep-th]].
  M.~Dine, G.~Festuccia, A.~Morisse and K.~van den Broek,
  ``Metastable Domains of the Landscape,''
  JHEP {\bf 0806}, 014 (2008)
  [arXiv:0712.1397 [hep-th]].
  P.~Narayan and S.~P.~Trivedi,
  ``On The Stability Of Non-Supersymmetric AdS Vacua,''
  arXiv:1002.4498 [Unknown].
}

\lref\seibwit{
N.~Seiberg and E.~Witten,
  ``The D1/D5 system and singular CFT,''
  JHEP {\bf 9904}, 017 (1999)
  [arXiv:hep-th/9903224].
}

\lref\hcc{
T.~Hertog and G.~T.~Horowitz,
  ``Holographic description of AdS cosmologies,''
  JHEP {\bf 0504}, 005 (2005)
  [arXiv:hep-th/0503071].}

  \lref\alberghi{
  G.~L.~Alberghi, D.~A.~Lowe and M.~Trodden,
  ``Charged false vacuum bubbles and the AdS/CFT correspondence,''
  JHEP {\bf 9907}, 020 (1999)
  [arXiv:hep-th/9906047].
  }
  
  \lref\lowe{
David A. Lowe, ``Some comments on embedding inflation in the AdS/CFT correspondence,"
Phys.\ Rev. \ D{\bf 77}, 066003 (2008) 
 arXiv:0710.3564 [hep-th].
}
  
\lref\dtrace{
E.~Witten,
  ``Multi-trace operators, boundary conditions, and AdS/CFT correspondence,''
  arXiv:hep-th/0112258.
  M.~Berkooz, A.~Sever and A.~Shomer,
  ``Double-trace deformations, boundary conditions and spacetime
  singularities,''
  JHEP {\bf 0205}, 034 (2002)
  [arXiv:hep-th/0112264].
}

\lref\fubini{
S.~Fubini, ``A New Approach To Conformal Invariant Field Theories,''
Nuovo Cim.\ A {\bf 34}, 521 (1976).}

\lref\zinn{
J.~Zinn-Justin, ``Quantum Field Theory and Critical Phenomena", Clarendon Press, Oxford 1996.}

\lref\linde{
A.~D.~Linde,
  ``Particle Physics and Inflationary Cosmology,''
  arXiv:hep-th/0503203.
}
\lref\rabroll{
V.~Asnin, E.~Rabinovici and M.~Smolkin,
  ``On rolling, tunneling and decaying in some large N vector models,''
  JHEP {\bf 0908}, 001 (2009)
  [arXiv:0905.3526 [hep-th]].
  }
  
  \lref\marga{
  J.~Garcia-Bellido, M.~Garcia Perez and A.~Gonzalez-Arroyo,
  ``Symmetry breaking and false vacuum decay after hybrid inflation,''
  Phys.\ Rev.\  D {\bf 67}, 103501 (2003)
  [arXiv:hep-ph/0208228].
}

\lref\craps{
A.~Bernamonti and B.~Craps,
  ``D-Brane Potentials from Multi-Trace Deformations in AdS/CFT,''
  JHEP {\bf 0908}, 112 (2009)
  [arXiv:0907.0889 [hep-th]].
}

\lref\horo{
T.~Hertog and G.~T.~Horowitz,
  ``Towards a big crunch dual,''
  JHEP {\bf 0407}, 073 (2004)
  [arXiv:hep-th/0406134].
}

\lref\eli{
S.~Elitzur, A.~Giveon, M.~Porrati and E.~Rabinovici,
  ``Multitrace deformations of vector and adjoint theories and their
  holographic duals,''
  JHEP {\bf 0602}, 006 (2006)
  [arXiv:hep-th/0511061].  {\it ibid} Nucl.\ Phys.\ Proc.\ Suppl.\  {\bf 171}, 231 (2007).
  
}

\lref\dtrace{
E.~Witten,
  ``Multi-trace operators, boundary conditions, and AdS/CFT correspondence,''
  arXiv:hep-th/0112258.
  M.~Berkooz, A.~Sever and A.~Shomer,
  ``Double-trace deformations, boundary conditions and spacetime
  singularities,''
  JHEP {\bf 0205}, 034 (2002)
  [arXiv:hep-th/0112264].
  }
  
\lref\ofer{
O.~Aharony, B.~Kol and S.~Yankielowicz,
  ``On exactly marginal deformations of N = 4 SYM and type IIB  supergravity on
  AdS(5) x S**5,''
  JHEP {\bf 0206}, 039 (2002)
  [arXiv:hep-th/0205090].
}

\lref\turok{
 B.~Craps, T.~Hertog and N.~Turok,
  ``Quantum Resolution of Cosmological Singularities using AdS/CFT,''
  arXiv:0712.4180 [hep-th].
 
  N.~Turok, B.~Craps and T.~Hertog,
  ``From Big Crunch to Big Bang with AdS/CFT,''
  arXiv:0711.1824 [hep-th].
}

\lref\affleck{
I.~Affleck,
  ``Quantum Statistical Metastability,''
  Phys.\ Rev.\ Lett.\  {\bf 46}, 388 (1981).
}

\lref\roberto{
M.~M.~Caldarelli, O.~J.~C.~Dias, R.~Emparan and D.~Klemm,
  ``Black Holes as Lumps of Fluid,''
  JHEP {\bf 0904}, 024 (2009)
  [arXiv:0811.2381 [hep-th]].
  R.~Emparan, T.~Harmark, V.~Niarchos and N.~A.~Obers,
  ``Blackfolds,''
  Phys.\ Rev.\ Lett.\  {\bf 102}, 191301 (2009)
  [arXiv:0902.0427 [hep-th]];
  ``Essentials of Blackfold Dynamics,''
  arXiv:0910.1601 [Unknown];
  ``New Horizons for Black Holes and Branes,''
  arXiv:0912.2352 [Unknown].
}

\lref\polhorow{
G.~T.~Horowitz, J.~Orgera and J.~Polchinski,
  ``Nonperturbative Instability of AdS$_5$ x S$^5/Z_k$,''
  Phys.\ Rev.\  D {\bf 77}, 024004 (2008)
  [arXiv:0709.4262 [hep-th]].
}

\lref\bekus{
J.~L.~F.~Barbon and E.~Rabinovici,
  ``Closed-string tachyons and the Hagedorn transition in AdS space,''
  JHEP {\bf 0203}, 057 (2002)
  [arXiv:hep-th/0112173];
  ``Remarks on black hole instabilities and closed string tachyons,''
  Found.\ Phys.\  {\bf 33}, 145 (2003)
  [arXiv:hep-th/0211212].
  }
  
\lref\sussk{
 L.~Susskind,
  ``The anthropic landscape of string theory,''
  arXiv:hep-th/0302219.
  }
  
  \lref\rbek{ J.~L.~F.~Barbon and E.~Rabinovici,
  ``Remarks on black hole instabilities and closed string tachyons,''
  Found.\ Phys.\  {\bf 33}, 145 (2003)
  [arXiv:hep-th/0211212].
  }
  
\lref\tassos{
S.~de Haro and A.~C.~Petkou,
  ``Instantons and conformal holography,''
  JHEP {\bf 0612}, 076 (2006)
  [arXiv:hep-th/0606276].
  }
 
 \lref\tpapa{
 S.~de Haro, I.~Papadimitriou and A.~C.~Petkou,
  ``Conformally coupled scalars, instantons and vacuum instability in
  AdS(4),''
  Phys.\ Rev.\ Lett.\  {\bf 98}, 231601 (2007)
  [arXiv:hep-th/0611315].
  }
 
 \lref\papa{
 I.~Papadimitriou,
  ``Multi-Trace Deformations in AdS/CFT: Exploring the Vacuum Structure of
  the Deformed CFT,''
  JHEP {\bf 0705}, 075 (2007)
  [arXiv:hep-th/0703152].
  }


\baselineskip=15pt

\line{\hfill IFT UAM/CSIC-10-13}

\vskip 0.7cm

\Title{\vbox{\baselineskip 12pt\hbox{}
 }}
{\vbox {\centerline{Holography of   }
\vskip10pt
\centerline{ AdS}
\vskip10pt
\centerline{vacuum bubbles}}}

\vskip 0.5cm

\centerline{$\quad$ {\caps
Jos\'e L.F. Barb\'on$^\dagger$
 and
Eliezer Rabinovici$^\star$
}}
\vskip0.7cm

\centerline{{\sl  $^\dagger$ Instituto de F\'{\i}sica Te\'orica IFT UAM/CSIC }}
\centerline{{\sl  C-XVI,
 UAM, Cantoblanco 28049. Madrid, Spain }}
\centerline{{\tt jose.barbon@uam.es}}

\vskip0.2cm

\centerline{{\sl $^\star$
Racah Institute of Physics, The Hebrew University }}
\centerline{{\sl Jerusalem 91904, Israel}}
\centerline{{\tt eliezer@vms.huji.ac.il}}

\vskip0.7cm

\centerline{\bf ABSTRACT}

 \vskip 0.3cm

 \noindent

We consider the fate of AdS vacua connected by tunneling events. 
A precise holographic dual of thin-walled Coleman--de Luccia bounces is proposed in terms of Fubini instantons in an unstable CFT. This proposal is backed by several qualitative and
quantitative checks, including the precise calculation of the instanton action appearing in evaluating the decay rate.  Big crunches manifest themselves as time dependent processes
which reach the boundary of field space in a finite time. The infinite energy difference  involved is identified on the boundary and highlights the ill-defined nature of the bulk setup. We propose a qualitative scenario in which the crunch is resolved by stabilizing the CFT, so that all attempts at crunching  always end up shielded from the boundary by the formation of  black hole horizons. In all these well defined bulk processes the configurations have the same asymptotics and are finite energy excitations. 
 
\vskip 0.5cm

\Date{March 2010}

\vfill

\vskip 0.1cm




\baselineskip=15pt

\newsec{Introduction}

\noindent

The presence of certain types of singularities in classical gravity as well  as the presence and properties of black objects presents serious challenges to any contender for  a more complete description of general coordinate invariant systems. 
The uncovering of the  Quantum Field Theory (QFT) holographic duals of Anti-de Sitter (AdS) string backgrounds has been a big step 
in providing a non perturbative definition of string theory \refs\adscft.  In some cases  the more accessible properties of QFT could be used to test and verify if certain challenges are indeed resolved by string theory.
For example the potential black hole information paradox was articulated for the case of 
black holes in AdS. Known properties of QFT have shown that in this case the non perturbative
definition of string theory has removed the sting out of the argument for a paradox.

This tool has the potential of becoming a double edged sword as it may turn out that it can be used to demonstrate that  physics around a certain string background corresponds to an ill defined QFT.
As there is no court of appeal available after a non perturbative definition this would mean that 
string theory failed to resolve the challenge.  However before drawing drastic conclusions
one should verify that indeed string theory was under the obligation to resolve that particular challenge. 
Not all challenges need to be met. In particular if the string background is derived for a set
of repulsive branes there is no need to search for a static solution, if a bulk configuration has an
infinite amount of energy on a certain Cauchy surface there is neither need nor possibility to save it from forming a big crunch singularity. 

It is this latter issue of the big crunch on which we focus in this paper. The AdS/CFT correspondence is an arena which is both rather well defined and also well suited to
address aspects of the decay of metastable states which may lead to a  crunch.
We study how big crunches manifest themselves in the boundary holographic description and utilize the QFT knowledge to learn how these configurations can be inoculated and stabilized. This provides a class of configurations which may seem to be leading to a big crunch but do avoid it. In the process of studying this problem we are led to develop a dictionary between the physics of boundary QFTs which have metastable states and bulk theories of gravity which have potentially metastable states.
For some earlier discussions on similar issues see \refs{\horo, \hcc, \eli, \turok,  \craps}.

Coleman and de Luccia (CdL) have studied the decay of  (non supersymmetric) metastable scalar field states in the presence
of gravity \refs\cdl. They have set up  the general formalism and studied  decays to and from a zero cosmological constant   state. They have shown that in the thin-wall approximation decays from a potentially metastable zero cosmological constant  state  to a lower, negative energy  AdS state, may either not occur at all or may be on the verge of a big crunch. The first situation occurs when the surface  tension  of the  bubble wall driving the decay is large enough relative to the difference in vacuum energy between the two states. The second situation will generically result, not
in the decay into the lower AdS state,  but lead instead to a big crunch once small perturbations are admitted.

This result remains essentially unchanged when one considers the properties of a potential decay among two different negative energy AdS states. As in AdS space the volume and surface scale in the
same way as a function of a length scale, the parameters of the potential predetermine if the decay
via bubble formation and expansion is energetically possible.
The case when the decay is allowed leads essentially to a big crunch. Considering the difficulties to define a local concept
of energy in general coordinate invariant systems, it is not obvious how to decide upon the basis of a bulk calculation alone  if a big crunch is acceptable or needs to be resolved. An infinite amount of energy on a given Cauchy surface should lead to an acceptable big crunch. It is in terms of the boundary CFT that the question is better posed and if the big crunch in the bulk is acceptable then one would expect that identifying its QFT dual would reveal an ill defined, incurable system.
We indeed uncover the dual QFT of the CdL bubble and it shows an unbounded potential. Moreover the
metastable state reaches  an infinite distance in field space in a finite time. 

Our results are obtained after rewriting the standard defect treatment \refs{\israel, \guthblau, \tkachev, \aurilia, \alberghi, \shenki}\  of thin-walled 
bubbles in terms of a theory of branes, defined  by effective tension and charge parameters. This 
presentation of the thin-wall approximation to the bubble dynamics has the advantage of 
suggesting the appropriate dual descriptions in an AdS/CFT framework. As a concrete 
result, we present an explicit duality relationship between spherical CdL bubbles in AdS 
and non-gravitational bubbles that mediate the decay through a barrier which is unbounded from below and conformal. 
This duality goes as far as matching exactly the instanton action for Fubini-type  
configurations in a critical scalar field theory \refs\fubini\ and the WKB exponent for the nucleation probability in the brane description of bubble dynamics, for a certain range of the parameters. This  calculation exhibits a remarkable matching of  non
supersymmetric quantities.

The next stage is to stabilize the system on the duality side which is  most transparent, i.e in the QFT framework.
We suggest how to do so and this results in a QFT which has  stable and metastable vacuum states.
This system itself is the subject of the original Coleman's treatment of the fate of a  false vacuum in a QFT.  We  suggest a system where  a potential but finite energy big crunch is indeed stopped on its tracks. The dual bulk configuration has hybrid features of a  CdL bubble whose expansion is interrupted and of a 
domain wall.  

This result has bearing on the issues raised in particular by Banks \refs\banks.
Our regularized analysis incorporates a natural feature of any finite-energy QFT state, 
namely that it looks like the vacuum when probed at very high energies. According to the 
UV/IR relation of AdS/CFT, this in turn implies that both  bulk conÞgurations 
must be identical Ônear the boundaryÕ of AdS. In particular, they must share the same value 
of the asymptotic cosmological constant. Hence, the standard CdL transitions with a lower 
vacuum energy bubble reaching the boundary cannot  correspond to quantum transitions 
with finite energy exchanges in the QFT.  In fact, the true 
vacuum configuration of any regularized decay also has the asymptotics of the vacuum corresponding to the higher energy AdS vacuum, although it  has transient features of the lower  AdS geometry in a finite region of bulk spacetime. The extent of that region is determined by the regulator.
The endpoint of the configuration formed  in its interior is a black hole which has swept under the protection of its horizon all the evidence of the potential crunch.

The paper is organized as follows. In section 2 we review the defect approach to thin-walled bubbles in AdS spacetimes, and rephrase it in terms of a brane-type action for the 
shells representing the walls. In section 3 we show how the formalism based on brane 
dynamics can be used to compute decay rates in thermally excited situations in either of 
the two vacua separated by the bubble wall. In section 4 we match these results to an 
effective  Conformal Field Theory (CFT) Lagrangian model, and present our statement regarding the dual of 
CdL bubbles. In section 5 we study the regularization of the ensuing bulk crunches by a 
stable UV potential in the CFT, and discuss the bulk interpretation of the decay endpoints 
resulting from this regularization of the problem.  We also suggest a very schematic 5-d  effective potential appropriate for these considerations.  We conclude by section 6 and discuss some technical issues in an Appendix.

\newsec{Thin-walled bubbles in AdS}

\noindent

Let us consider the simplest possible model of a `landscape' in the form of  a single  scalar field with potential $U(\chi)$, coupled to gravity   in $d+1 > 3$  spacetime dimensions. The potential is assumed to have two locally stable vacua  of negative energy densities
$U_- <U_+ <0$. The corresponding Anti--de Sitter   (AdS) cosmological constants in $d+1$ spacetime dimensions, $\lambda_\pm = -1/R_\pm^2$,  are related to the energy densities by  
$$
U_\pm =   {d(d-1) \lambda_\pm \over  16\pi G} \;,
$$
 where 
 $R_\pm$ are the curvature radii of the two AdS's. The vacuum closer to zero energy density, $\chi_+$, 
corresponds to an AdS `closer' to flat space over a larger distance scale.

We can introduce a certain amount of excitation in the system by adding Schwarzschild terms
of effective mass parameters  
\eqn\mus{
\mu_\pm = {16\pi G M_\pm  \over (d-1) v}\;,
}
where $v= |{\bf S}^{d-1}|$ denotes the volume of the angular $(d-1)$-dimensional sphere. Hence, the  ${\rm AdS}_\pm$ asymptotic metrics in $d+1$ spacetime dimensions 
read (in global coordinates)
\eqn\metrics{
ds_\pm^2 = -f_\pm (r) dt_\pm^2 + {dr^2 \over f_\pm (r)} + r^2 \,d\Omega_{d-1}^2
\;,}
with
\eqn\profi{
f_\pm (r) = 1 - \lambda_\pm \,r^2 - {\mu_\pm \over r^{d-2}}
\;.}

According to the basic rules of AdS/CFT  (cf. \refs\adscft), a global  AdS background admits  a dual nonperturbative  description in terms of a CFT with an effective number of degrees of freedom $N_{\rm eff} \sim R^{d-1} /G$,  where $R$ is the curvature radius of AdS. The CFT is  defined on the manifold ${\bf R} \times {\bf S}^{d-1}$, with the first factor representing the time coordinate, and the spatial sphere of radius $R$ introducing a spectral gap of order $1/R$. Hence, the model $U(\chi)$ provides us with two candidate quantum systems, associated to the two conformal field theories dual to AdS$_\pm$. 

 \bigskip
\centerline{\epsfxsize=0.4\hsize\epsfbox{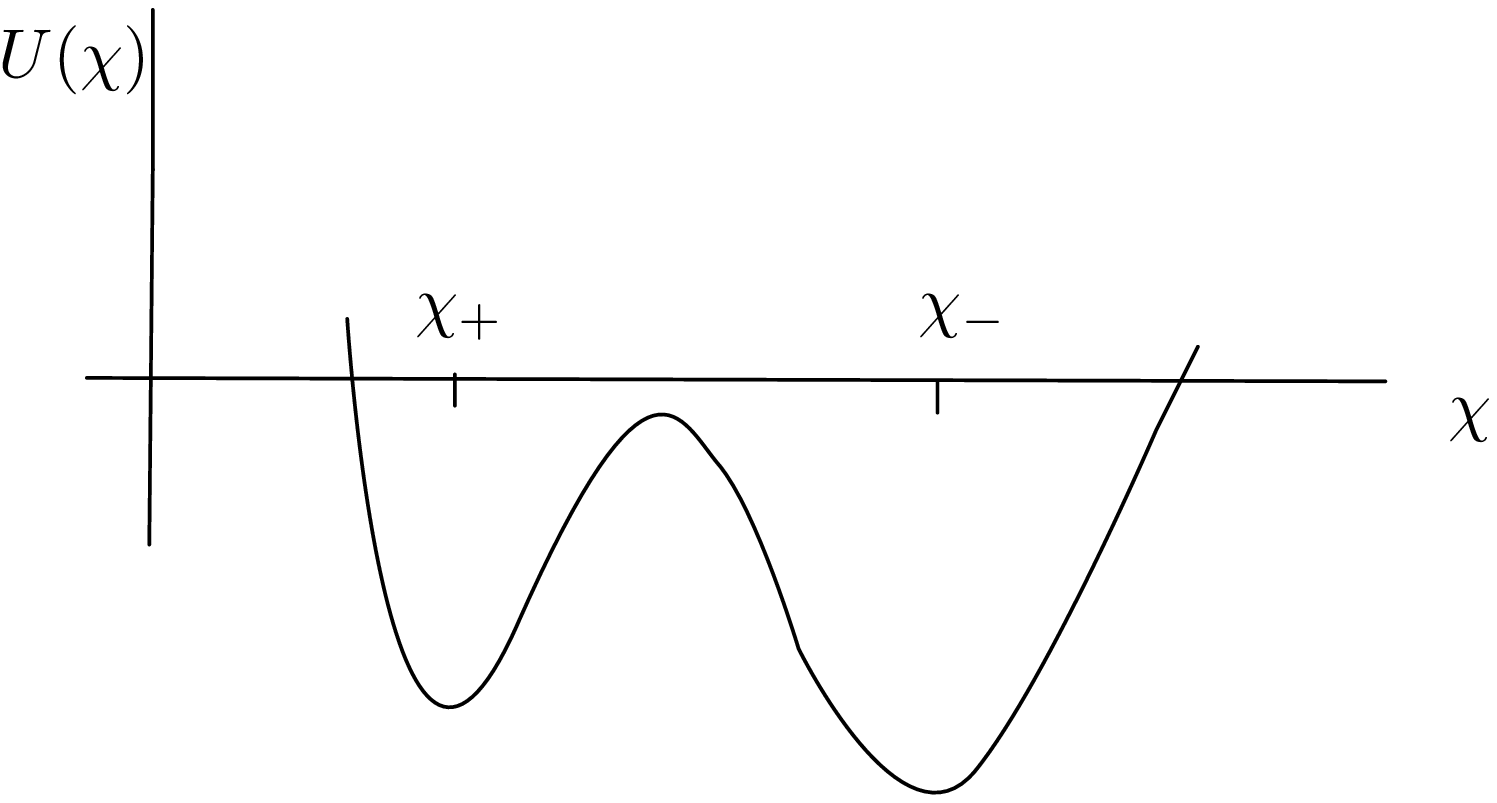}}
\noindent{\ninepoint\sl \baselineskip=8pt {\bf Figure 1:} {\ninerm
A scalar field model that reproduces two classical AdS vacua with negative cosmological constant proportional to $U_\pm = U(\chi_\pm) <0$. }}
\bigskip

In the thin-wall approximation, one may consider configurations of  bubbles of one metric immersed on the other, separated by a spherical wall whose thickness is negligible compared to its size. 
A bubble of the high vacuum (+) leaving the low vacuum $(-)$ outside is the situation studied by Guth and Fahri (GF) in \refs\rgf. The Coleman--de Luccia  (CdL) situation corresponds to a bubble of the low vacuum $(-)$ surrounded by the high vacuum (+) outside it.  

The total ADM mass of the CdL bubble, measured with respect to the AdS$_+$ vacuum, is given by $M_+$. If the interior of the bubble is an AdS$_-$ space with some internal $M_-$ excitation (such as a black hole at the center), 
then we refer to  the ADM mass of the shell forming the bubble wall, as $\omega = M_+ - M_-$, a quantity which may be either positive or negative. On the other hand, we shall restrict  the AdS$_\pm$ mass parameters to be non-negative,  $M_\pm \geq 0$ so that no naked singularities occur in the geometry.

The conditions for the quantum nucleation of an AdS$_-$ bubble inside an AdS$_+$ vacuum were studied in the original reference \refs\cdl\  applying the thin-wall approximation  to the $\chi$  field theory coupled to gravity. Here, we shall follow the analysis of  \refs{\shenki} to describe the bubbles directly as defects in general relativity, whereby all the details of the potential function $U(\chi)$ are folded into two parameters, given by the surface tension of the bubble, and the difference $\Delta U = U_+ - U_-$ of vacuum energies.\foot{Notice that local maxima of $U(\chi)$ may still be good metastable vacua provided the effective mass gap controlled by $d^2 U /d\chi^2$ is below the BF bound \refs\bt.}
This approach has the advantage of including all nonlinear  relativistic effects in the motion of the bubbles, and the disadvantage that it is fundamentally tied to the `thin' aspect of the walls from the beginning.

The thin-wall approximation was  revealed, already in the original paper \refs\cdl,  as inadequate to understand the long-time evolution of the system, in particular it misses the generic occurrence of crunch cosmological singularities as a result of the
back-reaction from the detailed scalar field dynamics during the phase of bubble growth. Despite this caveat, we will show that the physical singularity associated to the crunch can be identified within the thin-wall approximation in AdS spaces and, when
combined with the AdS/CFT dictionary, conveniently interpreted in terms of the dual CFT. 

\subsec{Junction  dynamics}

\noindent

We describe the bubble wall as a shell characterized solely by the surface tension, i.e. the energy-momentum tensor is
\eqn\tshell{
T^{ab}\big|_{\rm shell} = -\sigma\,h^{ab}\;,
}
where $h_{ab}$ is the induced metric at the bubble wall and $\sigma$ is the surface tension. 

The energy of the bubble of radius $r$, measured with respect to the AdS$_+$ vacuum, is estimated in order of magnitude as built  from a surface term of order $\sigma\, r^{d-1}$ and a (negative) volume term of order $-\Delta U \,r^d = -(U_+ - U_-)\, r^d$, when the bubble is small ($r\ll R_-$)  and  of order $-\Delta U\, R_- \,r^{d-1} $
when the bubble is large in units of the AdS$_-$ radius.  This energy function has  a potential barrier which a zero-energy shell can tunnel through, emerging at the turning point 
$
 {\bar r} \sim  \sigma
/ \Delta U $ for $\sigma \ll \Delta U R_-$. In the opposite regime of large tension, $\sigma \gg \Delta U R_-$, the energy function is monotonically increasing and no tunneling occurs.  The low-tension turning point and  the tunneling action diverge  as $\sigma / \Delta U R_-$ approaches a critical value of $\CO(1)$.

In more detail, the  dynamics of the wall is determined by the junction conditions \refs{\israel} 
\eqn\junction{
\Delta K_{ab} - h_{ab} \,\Delta K = -8\pi G\,T_{ab} \big|_{\rm shell} \;,
}
where $\Delta K_{ab}$ is the jump in extrinsic curvature at the shell (exterior minus interior), and $\Delta K = h^{ab} \Delta K_{ab}$. Using the explicit form of the energy-momentum tensor \tshell\ we can rewrite this equation as
\eqn\otrajunc{
\Delta K^{a}_b = - \kappa \,\delta^a_b\;,
}
where we have defined the related tension parameter
$$
\kappa = {8\pi G \over d-1} \, \sigma\;.
$$
Computing the extrinsic curvature for the case of a spherical shell of induced metric
\eqn\metshell{
ds^2 \big|_{\rm shell} = -d\tau^2 + r(\tau)^2 \,d\Omega_{d-1}^2\;,
}
one finds
\eqn\roots{
\left[\left({dr \over d\tau}\right)^2 + f_- (r) \right]^{1\over 2} - 
\left[\left({dr \over d\tau}\right)^2 + f_+ (r)\right]^{1\over 2}  = \kappa\,r(\tau)\;
}
for  the equation determining the shell's trajectory $r(\tau)$ as a function of its proper time. Squaring this equation one can picture the dynamics in the form of a zero-energy motion in an  effective non-relativistic
potential problem \refs\shenki: 
\eqn\potu{
\left({dr \over d\tau}\right)^2 +{U}_{\rm eff} (r) =0
\;,}
 where 
\eqn\effpu{
{ U}_{\rm eff} (r) =
f_+ (r) - \left( {\kappa^2 r^2 + f_+ (r) - f_- (r) \over 2\kappa r}\right)^2
\;.}
Expanding out the terms in the potential we may rewrite  it in the form presented in ref. \refs\shenki, 
\eqn\exppo{
U_{\rm eff} (r) = 1-A\,r^2 - {B \over r^{d-2}} - {C \over r^{2d-2}}\;.
}
 The  coefficient $A$ controls the large $r$ behavior of the bubbles and  depends just on `vacuum' quantities,  $\lambda_\pm$ and $\kappa$. The coefficient $C$ is always positive and proportional to $(\Delta \mu)^2$, the squared mass of the shell, and controls the small $r$ motion of the bubbles, while $B$  depends linearly on the mass parameters $\mu_\pm$ and determines the transient behavior at intermediate radii. This potential problem describes the motion of thin-walled bubbles for any value of $\lambda_\pm$, including positive values appropriate for de Sitter type bubbles. Those were  studied extensively in \refs{\alberghi, \shenki, \lowe}.

In either GF or CdL cases, one usually thinks of a quantum nucleation of a bubble at radius ${\bar r}$
at $\tau=0$, and subsequent classical evolution past this point. Hence, the nucleated bubble is an initial condition for normal classical evolution for $\tau>0$. It is useful to consider situations in which the $\tau>0$ solution is reflected back to $\tau<0$ and thus look at time-symmetric solutions. The bubbles that come
out of tunneling correspond then to the turning points of the classical motion for time-symmetric solutions.

\vskip1cm
\centerline{\epsfxsize=0.4\hsize\epsfbox{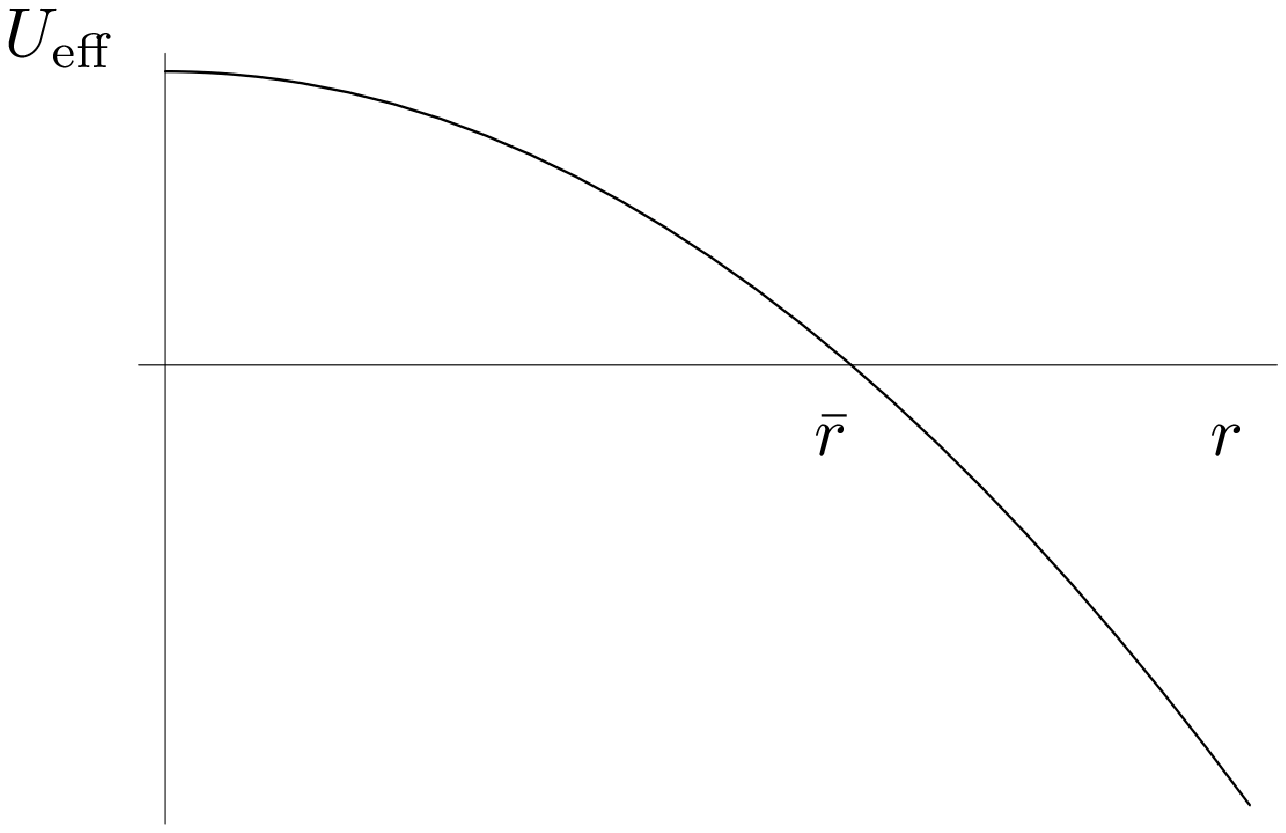}}
\noindent{\ninepoint\sl \baselineskip=8pt {\bf Figure 2:} {\ninerm
The proper-time  effective potential $U_{\rm eff} (r)$ in the zero-temperature CdL case, $\mu_\pm =0$. There is only one possible turning point at ${\bar r}$, corresponding to runaway trajectories, which diverges as $\kappa \rightarrow \kappa_c$ from below  (the potential approaching a straight line at $U_{\rm eff} =1$). }}
\vskip1cm

The CdL situation described above corresponds to $\mu_\pm =0$, leading to a
very simple quadratic potential, $U_{\rm eff} (r) = 1 -A \,r^2$. 
 There are turning points only in the case that $A>0$, and all trajectories come from $r=\infty$, hit the turning point at $\tau=0$ and then go off to infinity again (the CdL discussion would correspond to the $\tau\geq 0$ half of this). 
The
condition for $A>0$ is that $\kappa$ be {\it outside} the interval $[\kappa_c , \kappa_c']$ with
\eqn\kappas{
\kappa_c = {1\over R_-} -{1\over R_+}\;, \qquad \kappa_c' = {1\over R_-} + {1\over R_+}
\;.}
The turning point (equal to the nucleation size of CdL) satisfies
\eqn\tp{
{\bar r}^2 ={1\over A} = {4\kappa^2 \over (\kappa^2 - \kappa_c^2 )(\kappa^2 - \kappa_c'^{\,2})}
\;,}
so that ${\bar r} \rightarrow \infty$ as $\kappa \rightarrow \kappa_c$. This agrees with the CdL analysis.
However, the existence of turning points for $\kappa > \kappa_c'$, not anticipated in the original CdL analysis, is an unphysical feature of the equation \potu, which is obtained by squaring the
true equations of motion in order to get rid of nonlinear terms in  $dr/d\tau$. Filtering out the physical solutions of \potu\ requires the computation of the extrinsic curvature and the explicit verification of the correct signs in  equation  \junction\ (cf. \refs{\guthblau,\alberghi, \shenki}).  
The sorting of these subtleties, as well as a convenient starting point to discuss nucleation rates, is most conveniently done in a brane-action form of the shell dynamics, as we introduce in the next subsection.

\subsec{Bubble walls as branes}

\noindent

In order to relate the mechanics of the shell  to physical quantities defined in the dual QFT, it is  useful to recast the proper-time dependence into asymptotic time dependence,  $t_+$, because this is the time variable with a direct physical interpretation on the UV definition of the  QFT side. We thus  transform 
$$
{dr \over d\tau} = {dt \over d\tau} \,{\dot r}
$$
where ${\dot r} = dr/dt$. In what follows, we shall denote  $t=t_+$ and $f=f_+$, and we make a choice of units so that $R_+ =1$.  

We compute $dt/d\tau$ by  matching the exterior metric $ds^2_+$ in \metrics\  at the shell locus $r=r(\tau)$, to the induced metric at the shell \metshell, to find
\eqn\cambio{
{dt \over d\tau} = \sqrt{{1 \over f(r)} + {(dr/d\tau)^2 \over f(r)^2}}\;.
}
Plugging this expression back into \potu\ we find the equivalent potential problem in terms of
asymptotic time, expressed again as the  zero-energy motion 
\eqn\aspotu{
{\dot r}^2 + V_{\rm eff} (r) =0\;,
}
with
\eqn\potb{
V_{\rm eff} (r) = f(r)^2 \left[{f(r) \over f(r)-{U}_{\rm eff}(r)} - 1 \right]\;.
}
Using the  form \effpu\ of $U_{\rm eff}$ we give an explicit expression for  $V_{\rm eff}$ as
\eqn\vpot{
V_{\rm eff} (r) = f(r)^2 \left[{\sigma^2 \, v^2 \,r^{2d-2} \,f(r) \over (q\,v\, r^d + \omega)^2 } -1 \right]\;,
}
where we have defined
\eqn\defss{
q = (d-1)  {\Delta \lambda - \kappa^2 \over 16\pi G}\;, \qquad \Delta \lambda = \lambda_+ - \lambda_-\;,
}
and $\omega$ is the ADM mass of the shell, related to the difference in mass parameters $\Delta \mu = \mu_+ - \mu_-$ by the standard formula 
\eqn\omegam{
\omega = M_+ - M_- =  {(d-1)\, v \over 16\pi G} \Delta \mu
\;.}

The particular presentation of $V_{\rm eff}$ in \potb, based on the parameters $q, \sigma$ and $\omega$, finds its rationale  in the fact that \aspotu\   follows from a brane-type action of the form
\eqn\branea{
I = \int dt \,L = -\sigma \,v\,\int dt\,r^{d-1} \sqrt{f(r)- {{\dot r}^2 \over f(r)} } + q  \,v\, \int dt \,r^d\;,
}
where $\omega$ emerges naturally as  the canonical energy 
\eqn\canoe{
\omega = {\dot r} \,p_r - L = {\dot r} \,{\pt L \over \pt {\dot r} } - L 
}
associated with this Lagrangian, 
\eqn\formen{
\omega= {\sigma \,v\, r^{d-1} \, f(r)\over \sqrt{f(r)-{{\dot r}^2 \over f(r)}}} -q\,v\,r^d\;.
}
Squaring this equation and solving for ${\dot r}^{\,2}$ we readily obtain \vpot. 
 
The brane nature  of  \branea\  is rendered more explicit  when considering the spherical geometry of the shell and the induced metric, i.e. we can rewrite  \branea\ in the form  
\eqn\dbif{
I=-\sigma\,\int_{\cal W} \sqrt{-{\rm det}(h_{ab})} +q\,\int_{\cal W} C_d\;,
}
where ${\cal W}$ is the worldvolume of the shell, $h_{ab}$ its induced metric, and $C_d$ a $d$-form defined by
 \eqn\rrf{
C_d =    r^d \,dt \wedge dv\;,
}  
up to a closed form. \foot{The ambiguity by a closed form translates into boundary terms in the action \dbif, such as different additive normalizations of the energy $\omega$. The conventions adopted here are those that ensure the ADM relation $\omega = M_+ - M_-$.} The exterior derivative of $C_d$  is proportional to the volume form of AdS,  $dC_d = (d)\,r^{d-1}\,dr\wedge dt \wedge dv$, with $dv$ the volume for of ${\bf S}^{d-1}$. In this way we can rewrite the effective charge coupling in
\dbif\ as a volume integral over a $(d+1)$-dimensional manifold having ${\cal W}$ as one boundary component, i.e. the charge coupling is of Wess--Zumino type.\foot{For a similar incarnation of effective branes in more general situations of strong gravitational dynamics see \roberto.}

It is interesting to dissect the form of the effective brane charge $q$, by writing it in terms of primitive dynamical quantities 
\eqn\dissect{
q= q_0 - {4\pi G \over d-1} \,\sigma^2\;,
}
where
\eqn\qcero{
  q_0 = (d-1) {\Delta \lambda \over 16\pi G} = {\Delta U \over d}\;,
}
is the effective charge in the absence of gravitation, i.e. in the $G\rightarrow 0$ limit.  It  is proportional to the vacuum energy difference, elucidating  the nature of the Wess--Zumino term in \dbif\ as a purely volume contribution to the energy. It is however  peculiar to find that $q_0$ is renormalized additively by a `surface' term, proportional to $\sigma^2$. In fact, this term can be interpreted as a `surface binding energy' of the shell, and always has a volume scaling. In Newtonian terms, the effective mass of the shell is $\sigma v r^{d-1}$, and the associated gravitational self energy is  proportional to $G (\sigma v r^{d-1})^2 / r^{d-2} \sim G\sigma^2 r^d$, thus mimicking a volume term.

\subsec{Qualitative dynamics}

\noindent

In going from \branea\ to \aspotu\ and \vpot\ we must take the square of \canoe. In this process, some information about signs is lost, so that the set of solutions to \aspotu\ is actually larger than the physical set of trajectories determined by \branea. We can resolve the ambiguity by looking
at the explicit form of \canoe:
\eqn\canoee{
\omega + q \,v\,r^d = {\sigma \,v\, r^{d-1} \, f(r) \over \sqrt{f(r)-{{\dot r}^2 \over f(r)}}}\;.
}
The positivity of the right hand side translates into the following rule: given the location of the pole of \vpot, $r_\omega = (- \omega/qv)^{1/d}$, we can only find  positive charge branes ($q>0$) propagating to the right of the pole, and 
 negative-charge branes ($q<0$) propagating to the left of the pole. 
 In terms of bulk variables the constraint $\omega + q\,v\,r^d >0$ is equivalent to  the inequality
\eqn\excrule{
(\Delta \lambda - \kappa^2) \,r + {\Delta \mu \over r^{d-1}} >0\;,
}
which in particular implies that only branes with $\Delta \lambda > \kappa^2$ can propagate to asymptotically large radii.  Since we are
precisely interested in CdL bubbles that are able to reach the boundary, we henceforth concentrate on the case
$\Delta \lambda > \kappa^2$ in what follows. 

Shells bounding bubbles of the GF type, with AdS$_-$ on the asymptotic region, will be called {\it antibranes}.   Their effective potential with respect to the asymptotic time, $t_-$ in this case, is obtained from \vpot\ by setting $f=f_-$ and switching the $+$ and $-$ labels, which in turn implies that the effective charge is negative for antibranes. As a consequence, no bubble leaving  AdS$_-$ in its exterior can ever hit $r=\infty$. \foot{This fact would pull the rug under any attempt one might consider of sending antibranes as  protective measures to collide
with runaway  branes stopping them on their way to crunch on the boundary.}

We may also use the {\it interior} time $t_+$ to describe the dynamics of the antibranes. In this case one finds
exactly the same potential \vpot\ as for branes, with effective values of  charge ${\bar q} = -q$ and canonical energy ${\bar \omega} =-\omega$. Again, we have the rule that branes propagate to the right of the pole, and antibranes do so to the left of the pole. Hence, we conclude that $V_{\rm eff} (r)$ in \vpot\ describes at the same time trajectoires of branes (in `exterior' time) and antibranes (for which $t_+$ is the  `interior' time).  

These considerations show that trajectories fall into different topological classes. We either have shells that bounce off a turning point and go to infinity, or we have shells that bounce off a turning point and go to smaller radii, falling into a black hole.  In this last case the bubble can be either of GF or CdL type (cf. figure 3).

 The effective potential $V_{\rm eff} (r)$ in \vpot\ describes bubbles of AdS$_-$ embedded in AdS$_+$, in terms of the exterior time $t_+$. 
 Since this coordinate is only defined outside horizons of the AdS$_+$ patch, we define the physical region for the trajectories described by $V_{\rm eff} (r)$  to be $r\geq r_0$, where $r_0 (M_+)$ stands for any  black hole horizon upon which the shell may impinge. For purely vacuum bubbles, the physical region is just $r\geq 0$.

\vskip1cm
\centerline{\epsfxsize=0.7\hsize\epsfbox{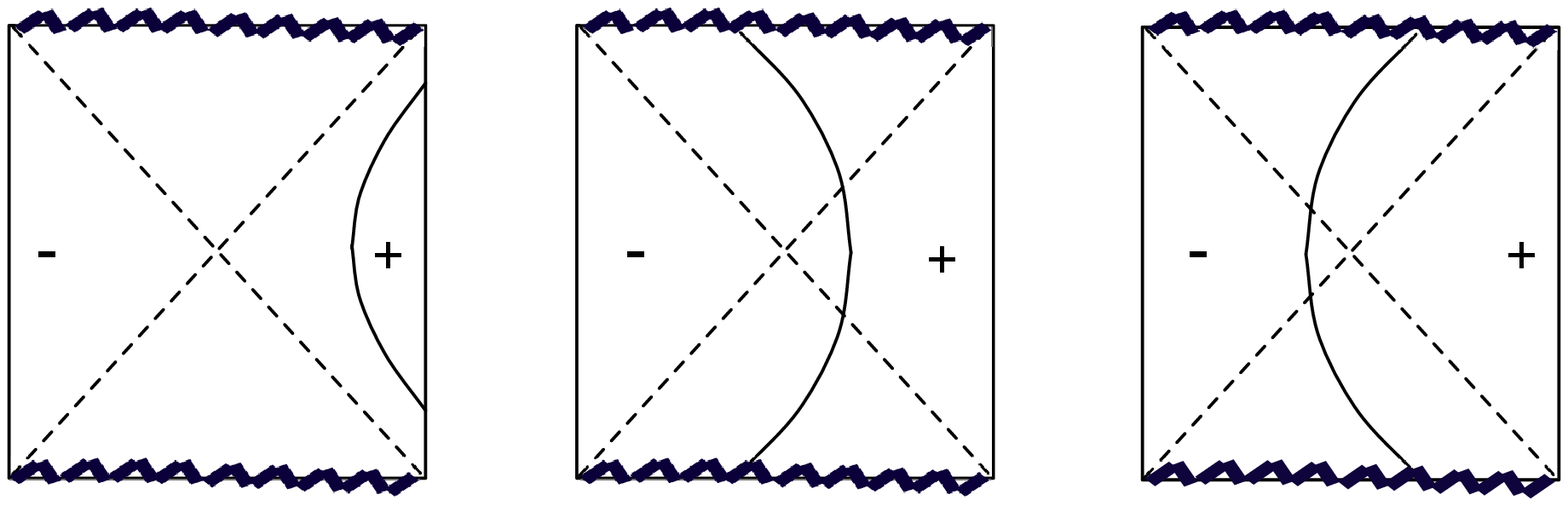}}
\noindent{\ninepoint\sl \baselineskip=8pt {\bf Figure 3:} {\ninerm  
The  basic time-symmetric bubble trajectories separating two AdSs with different curvature. To the right of the bubble we have here  AdS$_+$. The CdL and GF situations are interchanged by left-right reflection of the Penrose diagram. Notice that only those branes having  AdS$_+$ `on the outside'  may reach the boundary (CdL bubbles), whereas bubbles with a maximal radius can be of either CdL or GF type.}}
\vskip1cm

For shell energies $\omega > \omega_0 \equiv -q\,v\,r_0$ the pole in the potential $r_\omega = (-\omega/qv)^{1/d}$ is outside the physical region.  On the other hand, for $\omega < \omega_0$ the pole is located  in the physical region, $r_\omega > r_0$. In this case  the potential describes two types of objects.   Brane trajectories (bubbles of AdS$_-$ inside AdS$_+$)  take place in the asymptotic (large $r$) region   $r> r_\omega $, and antibrane trajectories  (bubbles of AdS$_+$ inside AdS$_-$) are restricted to the region `below' the pole, $r< r_\omega$. We can illustrate this situation by looking at figure 4, where the effective potential $V_{\rm eff} (r)$ is depicted for $M_+ = 0$, $M_- >0$ and $\omega = M_+ - M_- = -M_-  <0$. The zero-energy trajectories bouncing off the turning point ${\bar r}$ and going back to large radius correspond to branes, i.e. shells with vacuum AdS$_+$ outside and an AdS$_-$ black hole of mass $M_-$ inside. On the other hand, the trajectories occurring below the pole, for $r< {\bar r}'$, correspond to antibranes; shells with vacuum AdS$_+$ inside  and Schwarzschild AdS$_-$ outside, with mass parameter $M_- >0$. The ADM mass of this shell is ${\bar \omega} = M_- - M_+= -\omega >0$. Notice that this description, being the `interior' one, is oblivious to the black hole horizon, which can only be seen in an `exterior' description. In this case the exterior description of the antibrane would involve an effective potential with respect to $t_-$ time. 
  
The occurrence of poles in our brane effective potentials is a genuinely relativistic property, which can be found in more mundane situations, such as   an electron in a constant electric field, with Lagrangian
\eqn\schl{
L_e = -m_e\sqrt{1-{\dot x}^2} + e\,{\cal E}\,x\;.}
Performing the canonical analysis for this system we find an effective potential description
$
{\dot x}^2 + V_{\rm e} (x)=0\;,
$
with
\eqn\potel{
V_{\rm e} (x) = {m_e^2 \over (\omega + e\,{\cal E}\,x)^2} - 1\;,
}
with evident similarities to our brane potentials, including the rule that electrons propagate in the region $x>-\omega/ e{\cal E}$ and positrons do so in the region $x< -\omega /e{\cal E}$. The two turning points at $x_\pm = -\omega/e{\cal E} \pm m_e/e{\cal E}$ are interpreted semiclassically as the nucleation positions of  $e^+ e^-$ pairs in a  Schwinger decay process of the electric field.

By analogy with the $e^+ e^-$ situation we may contemplate `brane-antibrane' pairs obtained by a superposition of the  solutions discussed above. Namely a bubble of AdS$_+$ vacuum oscillating  inside a bubble of AdS$_-$, which is itself  immersed in the AdS$_+$ vacuum.  Such a configuration could arise if AdS$_+$ chose to decay by nucleating a `thick shell' made of AdS$_-$, rather than a spherical bubble of AdS$_-$. This configuration is topologically equivalent to a superposition of smaller AdS$_-$ bubbles disposed off center, making the AdS$_-$ shell. This suggests that these configurations are not dynamically important in themselves, since they can be constructed from the more elementary spherical bubbles. 

\vskip1cm
\centerline{\epsfxsize=0.4\hsize\epsfbox{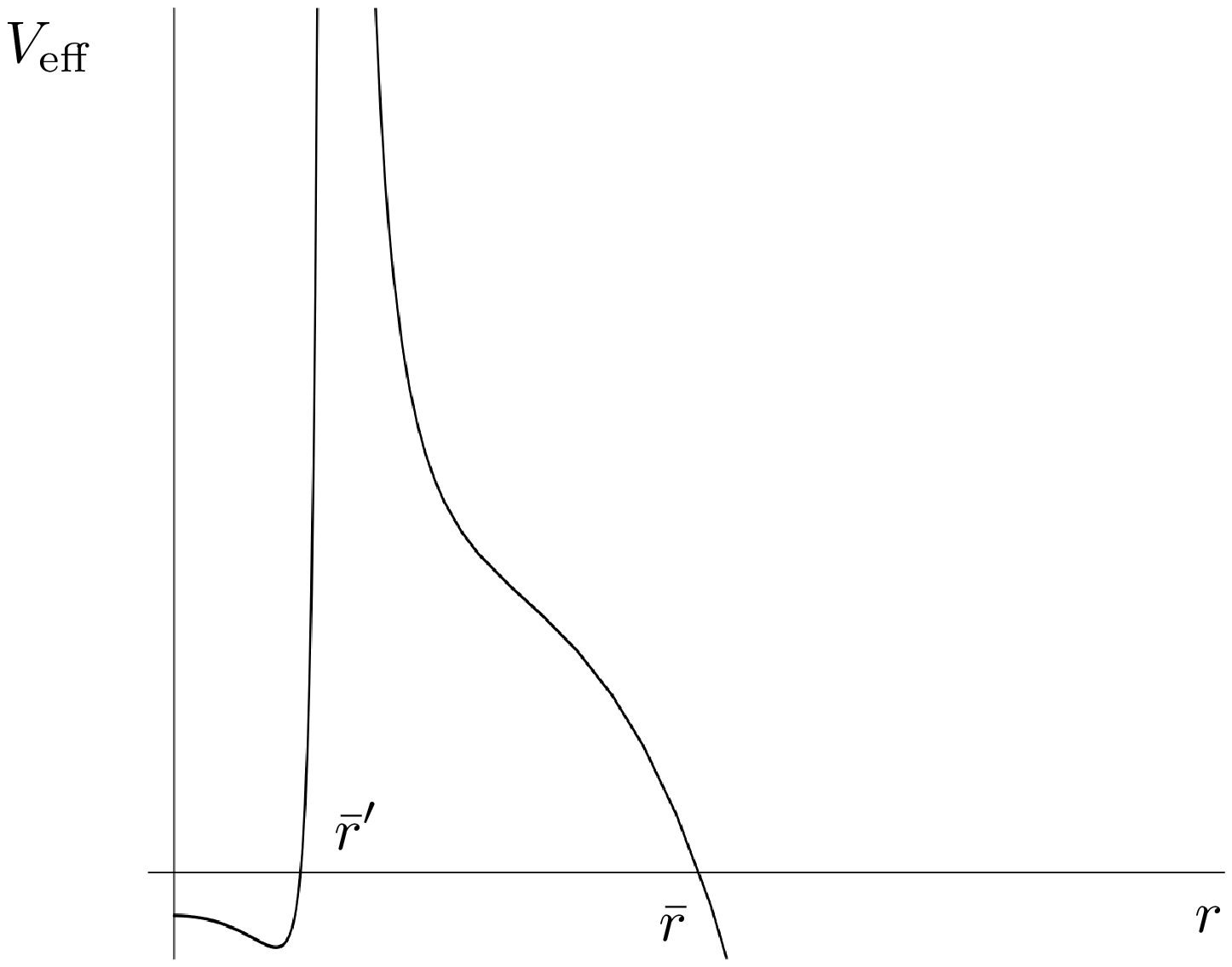}}
\noindent{\ninepoint\sl \baselineskip=8pt {\bf Figure 4:} {\ninerm
The effective potential describing the motion for $r> {\bar r}$ of vacuum bubbles of AdS$_-$ inside AdS$_+$, with parameters  $M_+ =0, M_- >0$ (the branes).    Solutions in the region $r<{\bar r}'$ describe bubbles of AdS$_+$ inside AdS$_-$ with the same parameters (the antibranes). The two types of solutions can be superimposed to describe concentric configurations of bubbles within bubbles. In all cases one describes the motions in terms of the static AdS$_+$ time variable.}}
\vskip1cm

 \subsec{ The CdL potential}
 
\noindent
 
 We can now go back to the basic CdL situation, corresponding to $\omega=0$  and $f(r) = 1+r^2$. The resulting potential is
\eqn\cdlp{
V_{\rm eff} (r)\big |_{\rm vac} = (1+r^2)^2 \left[ {\sigma^2 \over q^2} {1+r^2 \over r^2}  -1\right] \;.
}
It diverges as $r\rightarrow 0$ as $(\sigma/q)^2 /r^2$, and it is asymptotic to
\eqn\asympot{
\left[(\sigma/q)^2 -1\right] \,r^4 + 3\left[ (\sigma/q)^2 -1\right] \,r^2 + r^2
}
as $r\rightarrow \infty$. The last term, scaling as  $r^2$, is of geometrical origin. It is  proportional to the world-volume curvature of the brane, changing sign for negatively-curved world-volumes (see \javi\ for a study of this sort).   The condition for the existence of vacuum tunneling transitions
is that the tension be smaller than the charge in appropriate units, i.e. $\sigma < |q|$. We note that 
\eqn\notr{
{q^2 - \sigma^2 \over \sigma^2} = {(\kappa^2 - \kappa_c^2)(\kappa^2 - \kappa_c'^{\,2}) \over 4\kappa^2}\;,}
so that the $\sigma < |q|$ condition is equivalent to the previously stated condition that $\kappa$ be outside the interval $[\kappa_c , \kappa_c'\,]$.

When $\sigma = |q|$ we have 
a marginal situation where the leading terms in volume and surface energies cancel one another. In fact, for a flat world-volume, corresponding to $f(r)=r^2$, the potential vanishes altogether and we get back a familiar `no-force' condition on the brane signaling a situation of marginal stability. 

\vskip1cm
\centerline{\epsfxsize=0.4\hsize\epsfbox{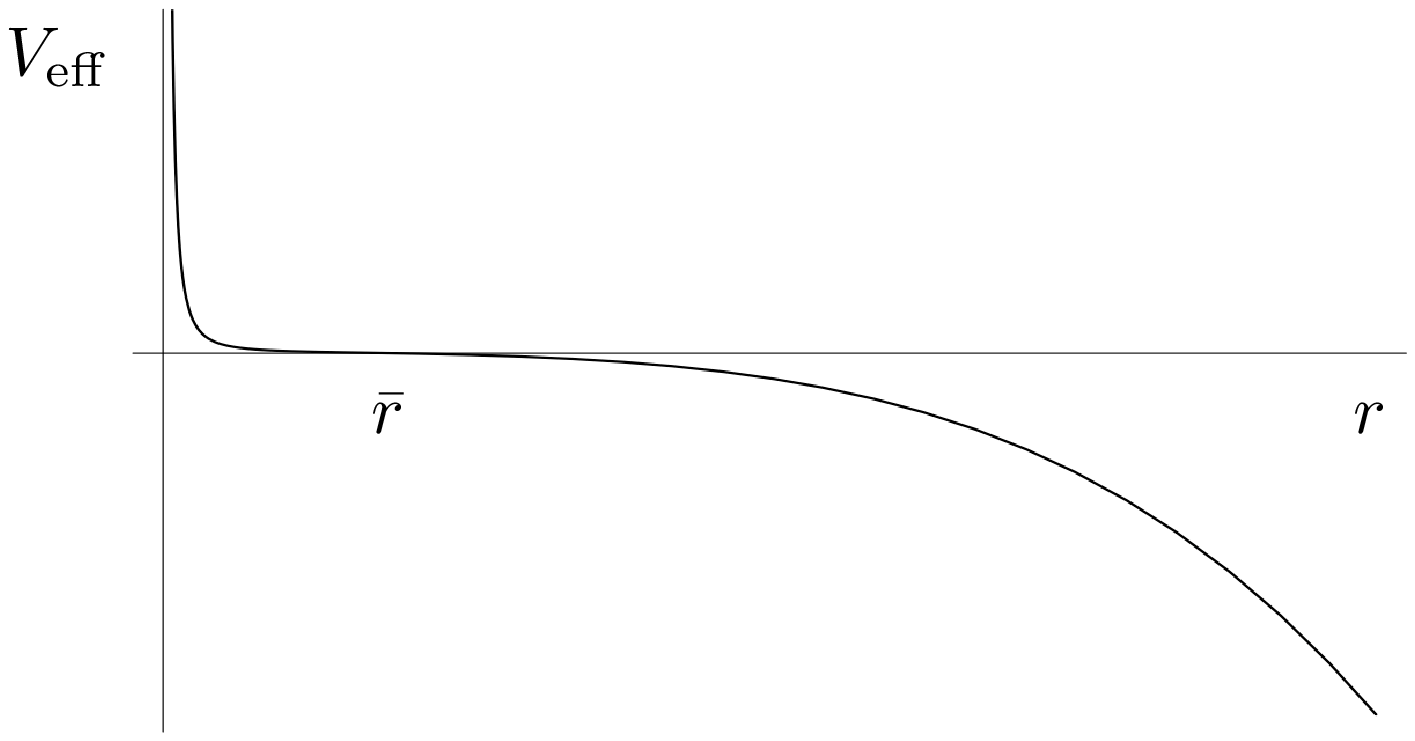}}
\noindent{\ninepoint\sl \baselineskip=8pt {\bf Figure 5:} {\ninerm
The asymptotic-time effective potential in the  CdL case, $f(r) = 1+r^2$ and $\omega =0$. }}
\vskip1cm

Restoring arbitrary units, we have $\sigma=|q|\,R_+$ which gives 
$$
\Delta U = {4\pi d\over d-1} \,G\,\sigma^2
$$
in the $R_+ \rightarrow \infty$ limit. This was identified in \cdl\ as the condition of marginal stability of Minkowski spacetime, and later interpreted as a supersymmetric relation  in the case that the theory admits a supersymmetric embedding (see  \refs{\susy, \rey, \susyr}). Hence, we interpret $|q| \leq \sigma$ as a BPS bound if the brane is to be understood as a (possibly nonsupersymmetric) excitation over a supersymmetric vacuum.

 We conclude that  nonperturbative instabilities
require violation of the BPS bound for the shell, with  implicit supersymmetry breakdown. In this case, the turning point, obtained from $V_{\rm eff} ({\bar r}) =0$, is given by
\eqn\tup{
{\bar r} = {\sigma /|q| \over \sqrt{1- \sigma^2 / q^2}} = {\alpha \over \sqrt{1-\alpha^2}}\;,
}
and indeed we find ${\bar r} \rightarrow \infty$ as $|q| \rightarrow \sigma$.  We denote $\alpha = \sigma/|q|$ the parameter characterizing the `degree of violation' of the bound.

From this one can find whether a running away bubble ever hits the boundary of AdS in finite asymptotic time, or `boundary QFT time'. The hit time from nucleation at ${\bar r}$ is
\eqn\hit{
t_{\rm hit} = \int_{\bar r}^\infty {dr \over {\dot r}} = \int_{\bar r}^\infty {dr \over \sqrt{-V_{\rm eff} (r)}} 
\;.
}
Using the asymptotics \asympot, we find $t_{\rm hit} \sim (1+{\bar r}^2 )/ {\bar r}$ in units of $R_+$. Hence,
the  bubble hits the boundary in finite time and in doing so its `kinetic energy', ${\dot r}^2$ diverges. \foot{Notice that the `proper hit  time' $\tau_{\rm hit}$ is infinite, because $U_{\rm eff} (r) \rightarrow -r^2 / {\bar r}^2$ as $r\rightarrow \infty$, a gentle fall.} This is a first  indication, within the thin wall approximation, of the existence of a singularity at $t_{\rm hit}$.  In fact, as already studied in the original CdL paper, the analysis beyond the thin-wall approximation reveals that the dynamics of scalar fields forces the interior of the bubble to crunch in finite time.

 \subsec{ Potential galore}
 
 \noindent
 
 We may now classify the qualitative form of the effective potentials as a function of the parameters. We shall restrict ourselves to the interesting case of $\alpha < 1$ and  to non-negative mass parameters $M_\pm \geq 0$ to avoid naked singularities. \foot{Potentials with $\alpha>1$ have no instabilities and the marginal BPS-saturated case, $\alpha=1$ is controlled by subleading curvature terms, cf. \javi.}This still leaves the mass of the shell $\omega = M_+ - M_-$ as a free parameter in the interval $[-\infty, M_+]$.

\vskip1cm
\centerline{\epsfxsize=0.4\hsize\epsfbox{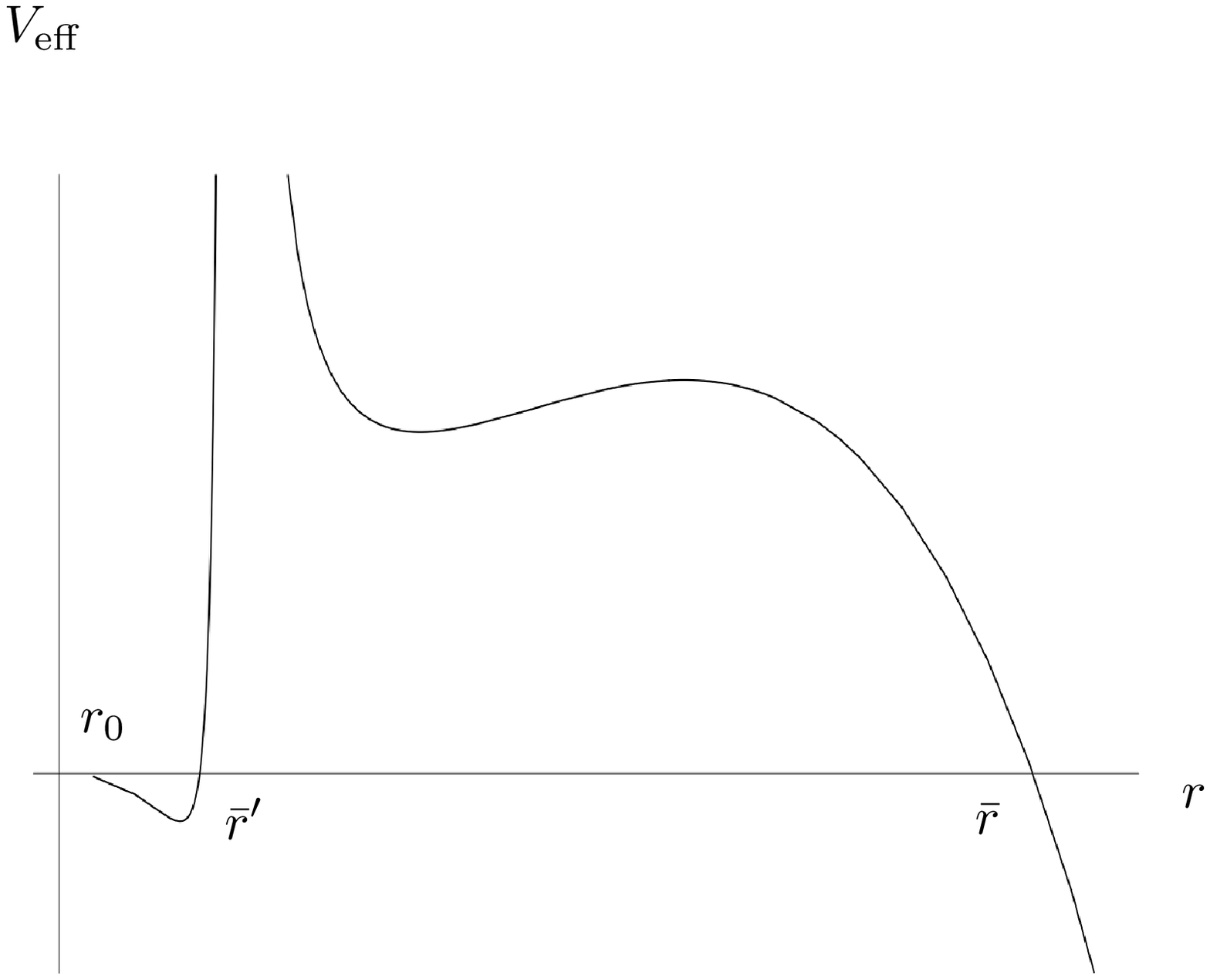}}
\noindent{\ninepoint\sl \baselineskip=8pt {\bf Figure 6:} {\ninerm
Plot of $V_{\rm eff} (r)$ for $\omega < \omega_0$, showing the pole in the physical region $r_\omega > r_0$. Trajectories in the interval $r_0 < r < {\bar r}'$ correspond to antibranes trapped near the horizon. Runaway trajectories in the region ${\bar r} < r< \infty$ correspond to branes. }}
\vskip1cm

 As mentioned above, shell energies above the critical value $\omega_0 = -q\,v\,r_0 (M_+)$ yield potentials without poles in the physical region. Conversely,  at very negative shell energies, $\omega < \omega_0$, the pole moves into the physical region. 
  For any $\omega \neq \omega_0$ we have $V_{\rm eff} (r_0) = V'_{\rm eff}(r_0) =0$, where the prime denotes derivative with respect to $r$. Furthermore the second derivative $V''_{\rm eff} (r_0) <0$, so that  
 the horizon is  a local maximum of the potential, which further vanishes  there. It follows that  the vicinity of the horizon is  an allowed region for brane (or antibrane) propagation. 
 For $\omega < \omega_0$ there are antibranes that oscillate between the horizon  and a turning point `below' the pole, i.e. in the region 
 $r_0 < r< {\bar r}' $ with ${\bar r}' < r_\omega$. Brane trajectories exist in the asymptotic region ${\bar r} < r<\infty$ (cf. figure 6).

 For the critical value of the energy, $\omega = \omega_0$, the potential still vanishes at the horizon, but the first derivative $V'_{\rm eff} (r_0) > 0$. Hence, there is a barrier extending from the horizon up to the large $r$ turning point, of order ${\bar r}$.

 For $\omega > \omega_0$ there are no poles in the physical region, and thus no antibrane trajectories either. Branes propagating in the vicinity of the horizon encounter a finite potential barrier for $\omega < \omega_s$, where $\omega_s$ is the `sphaleron' energy, for which the potential barrier degenerates to a single point.  
  For $\omega \gg \omega_s$ there is no barrier at all and the brane trajectories lay on the interval $r_0 < r< \infty$.  We summarize the cases $\omega \geq \omega_0$ in figure 7.

\vskip1cm
\centerline{\epsfxsize=0.4\hsize\epsfbox{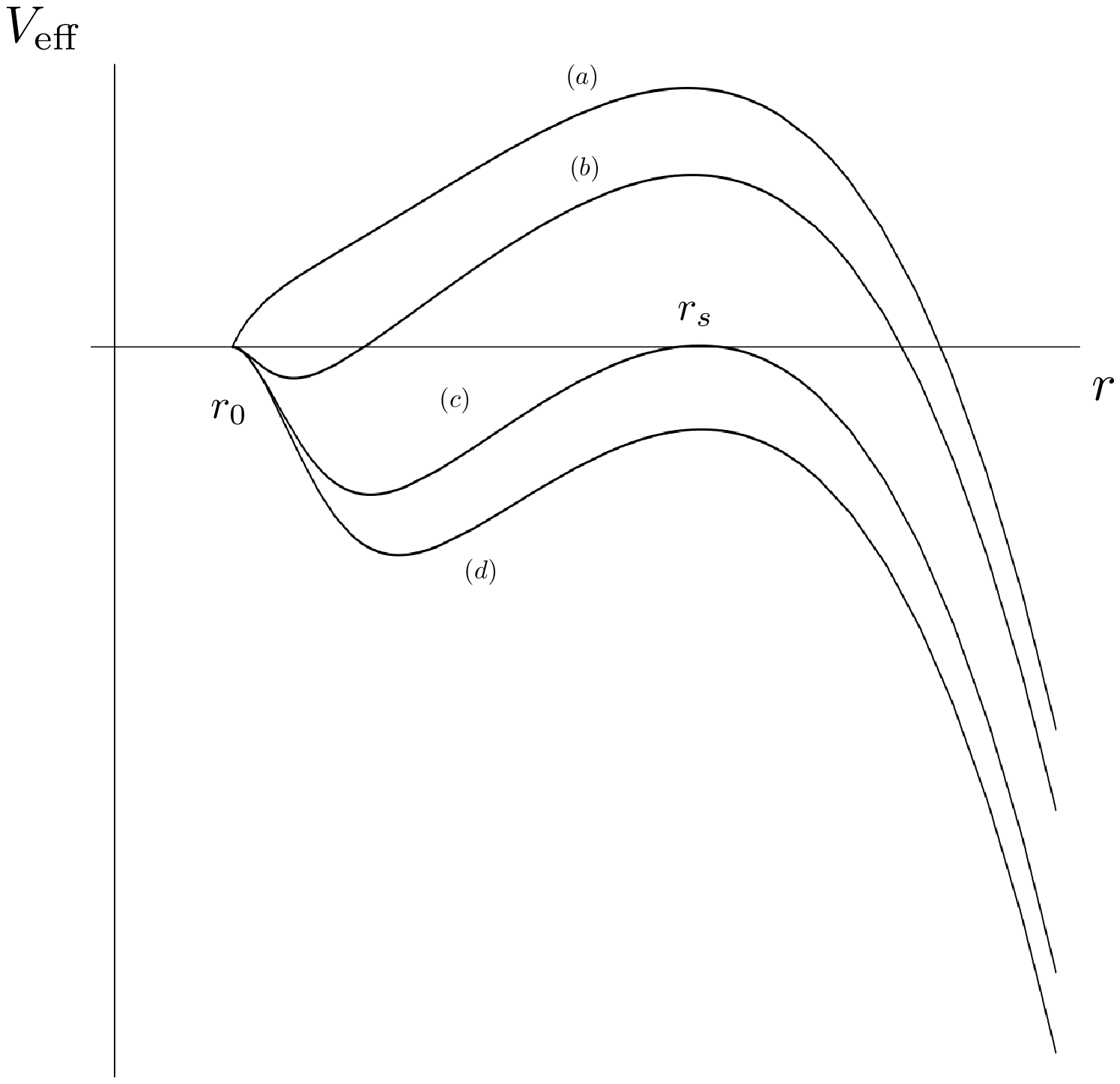}}
\noindent{\ninepoint\sl \baselineskip=8pt {\bf Figure 7:} {\ninerm
Plot of $V_{\rm eff} (r)$ with  four different  values of the shell energy. The upper curve $(a)$ corresponds to the critical value $\omega_a = \omega_0$ showing the barrier at the horizon. The curve $(b)$ has $\omega_0 < \omega_b < \omega_s$, where the barrier separates the brane trajectories bound to the horizon from the runaway trajectories. The curve $(c)$ corresponds to the `sphaleron' energy $\omega_c =\omega_s$, characterized by an unstable solution $r(t)=r_s$. Finally, for $\omega_d > \omega_s$ we have $(d)$, where branes emitted by the horizon propagate to infinity without further turning points. }}
\vskip1cm

\newsec{Decay rates}

\noindent

One  advantage of the brane picture for bubble nucleation is the existence of an action principle, in the form of \branea, which allows us to compute the rates for nucleation of spherical bubbles by using 
the standard quantum mechanical WKB approximation (see \tkachevq\ for a related but slightly different treatment). 

 The WKB ansatz for the wave function of a spherical brane  is of the form
\eqn\wkbwv{
\Psi_{\rm WKB} (t, r) \sim  \exp\left(-i\omega t + i\int^r p_{r'}\, dr' \right)
\;
,}
at energy $\omega$, 
 where $p_r = \pt L / \pt {\dot r}$, and the probability of barrier penetration in the leading exponential approximation is given by 
\eqn\probp{
P_{\rm WKB} \sim  \exp\left(-2\,{\rm Im}\, W(\omega)\right)\;,
}
where  
$
W= I + \omega t
$, with $I$ the action \branea. 
 The imaginary part in the classically forbidden region can be captured by the analytic continuation to the Euclidean signature $t=-it_E$ and we find 
 $
 {\rm Im}\, W(\omega) =W_E = I_E - \omega t_E \;,
 $ 
   with   $I_E$  the Euclidean action
\eqn\euclida{
I_E = \int dt_E \,L_E=  v\,\int d t_E \left[ \sigma\,r^{d-1} \sqrt{ f(r)+ {{\dot r}_E^2 \over f(r)}} -q\,  r^d  \right]\;.
}
 In terms of the canonical momentum 
$
(p_r)_E = {\pt L_E / \pt{{\dot r}_E}}
$
we can write 
$
W_E = \int dr \,(p_r)_E \;.
$
Finally, Euclidean trajectories correspond  to motion in the effective problem
\eqn\effeu{
{\dot r}_E^2 - V_{\rm eff} (r) =0\;,
}
which results form the real-time problem by a formal switch of the sign of the potential. 
Using this equation in the formula for the Euclidean action, we find the convenient expression
\eqn\tune{
2W_E (\omega) = 2\int_{\bar r'}^{\bar r} dr \,(p_r)_E= 2\,q\,v\,\int_{\bar r'}^{\bar r}  {dr  \over f(r)}\, \sqrt{\alpha^2 \,r^{2d-2} \,f(r) - (r^d - r_\omega^d)^2}\;,
} 
where the turning points in the integral are defined by the positivity of the square root argument, and $r_\omega^d = -\omega/qv$ is the location of the pole in the potential. 
This is a general formula for any value of the bubble parameters, including cases where the pole falls in the integration region, ${\bar r}' < r_\omega < {\bar r}$. The exponential WKB factor \tune\ is integrable across the pole and in this case it determines the rate of nucleation of brane/antibrane pairs, in a generalization of the Schwinger mechanism.\foot{For the $e^+ e^-$ potential \potel, equation \tune\ gives the expected exponent $2W_E =
2\int_{x_-}^{x_+} dx \sqrt{m_e^2 - (e{\cal E} x)^2} = \pi m_e^2/ e {\cal E}$ controlling the Schwinger effect amplitude.}

The total nucleation rate of shells  centered at $r=0$  takes the form
\eqn\totan{
\Gamma = \int_{\omega_{\rm min}}^{\omega_{\rm max}}  d\omega \,P_{\rm WKB} (\omega)\;,
}
where the limits in the integral correspond to the maximum and minimum possible energies of the shells. For the case $\alpha <1$, which is the subject of main interest here, the effective potential is unbounded from below and shells of arbitrarily negative energy can always propagate at sufficiently large radius, i.e. $\omega_{\rm min} =-\infty$. On the other hand, we require $M_\pm \geq 0$ in order to avoid a naked singularities in the interior of the AdS$_\pm$ bubbles, so that $\omega_{\rm max} = M_+$. 

\vskip1cm
\centerline{\epsfxsize=0.4\hsize\epsfbox{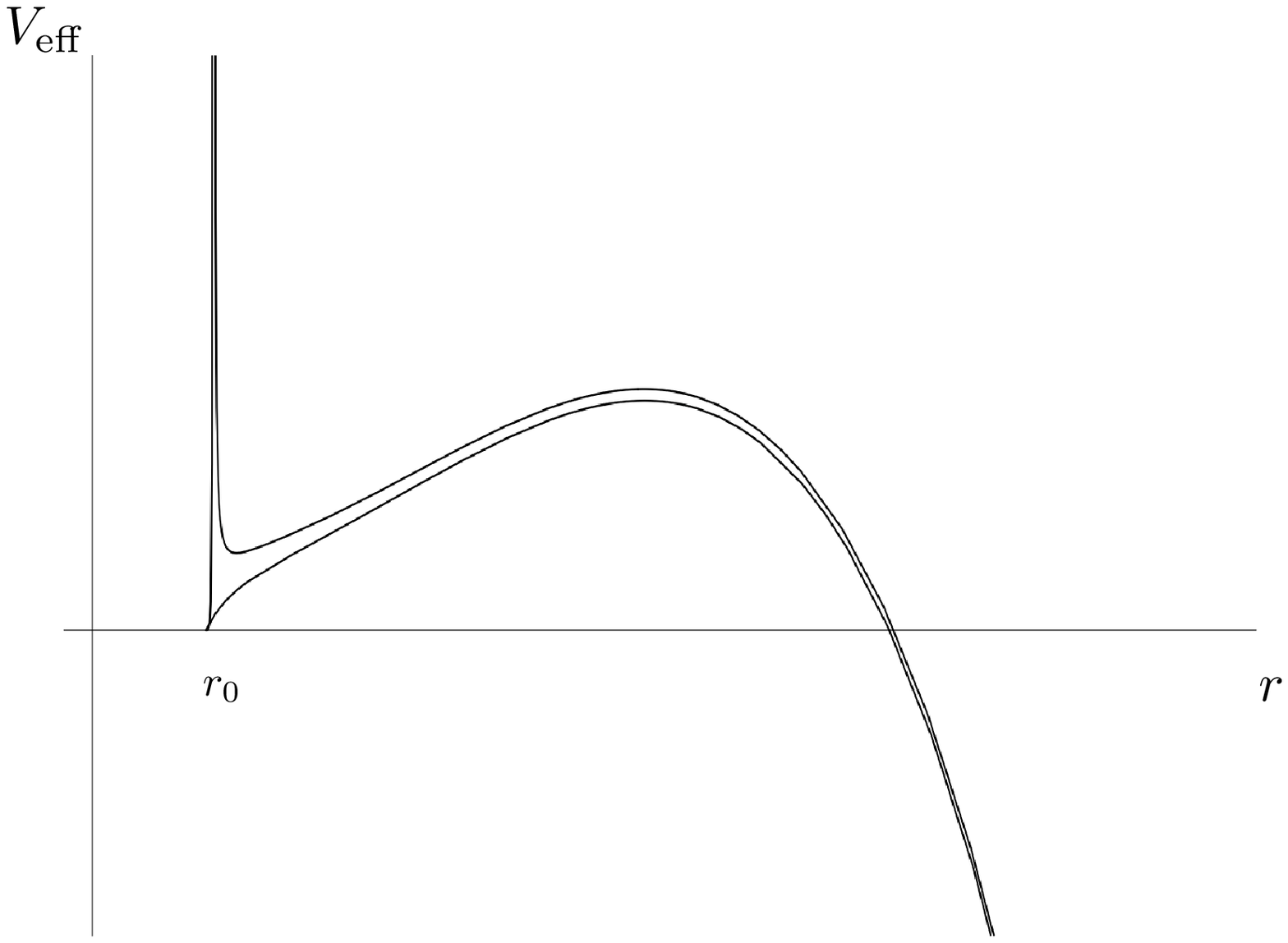}}
\noindent{\ninepoint\sl \baselineskip=8pt {\bf Figure 8:} {\ninerm
The effective potential for a shell energy slightly below the critical value $\omega<\omega_0$ (upper curve), showing the narrow pole
near the horizon which makes a very small correction to the $\omega=\omega_0$ barrier (lower curve). }}
\vskip1cm

Shells with  $\omega < \omega_0 = -q\,v\,r_0^d$ see an effective potential with a pole above the horizon. As stated before, this corresponds to Schwinger-like processes in which a concentric brane/antibrane pair is  nucleated above the horizon. For very negative values of $\omega$, the WKB exponent scales as  $2W_E  \propto qv(-\omega /qv)^{d-1 \over d}$, so that the nucleation of very large branes is suppressed. As $\omega$ approaches $\omega_0$ from below, the pole in the effective potential narrows down and makes a smaller and smaller contribution to the $\omega=\omega_0$ barrier (cf. figure 8). Hence, we conclude that the contribution to \totan\ coming from Schwinger-like processes is dominated by the endpoint and we may discard it when computing in the leading exponential approximation. 

The tunneling rate for pair production approaches  continuously that of single bubble nucleation. This is a reflection of the non-topological character of  spherical bubbles on global AdS, i.e. they can continuously shrink to zero size and disappear.

\subsec{Vacuum AdS decay}

\noindent

 In the CdL case, for a zero-mass bubble, $M_+ =0$,  bounded by a  zero-mass shell, $\omega=0$,  we have ${\bar r}' =0$  and $f(r) = 1+r^2$. The general expression for the tunneling exponent \tune\ reduces to  \foot{Notice that we take the initial tunneling condition at ${\bar r}' =0$ despite the pole at the origin of the zero-temperature   potential $V_{\rm eff} (r)$, since the barrier is still integrable. It can be regularized by introducing a small horizon radius, $r_0 > 0$ and taking the limit $r_0 \rightarrow 0$ at the end.}
\eqn\we{
2W_E = 2\,q\,v\, {{\bar r}^{d+1} \over \sqrt{1+{\bar r}^2}}  \int_0^{1} dx {x^{\,d-1} \over 1+{\bar r}^2 x^2} \sqrt{1 - x^2}\;.
}
This integral can be evaluated explicitly in terms of  beta and hypergeometric functions as
\eqn\exppp{
2W_E = 2 q\,v\,B(\sthreehalfs, \sdhalfs) \,{{\bar r}^{\,d+1} \over \sqrt{1+{\bar r}^{\,2}}} \,F\left[1,\sdhalfs, \sdmasthreehalfs, -{\bar r}^{\,2}\right]\;.
}
For small bubbles, ${\bar r} \ll 1$, we have
\eqn\smallb{
2W_E \Big |_{{\bar r} \ll 1}  \approx  q\,v\, B(\sthreehalfs, \sdhalfs)\,{\bar r}^{\,d+1}\;.
}
In this limit, corresponding to $\alpha \ll 1$, the charge parameter scales as 
$$
q\longrightarrow {(d-1)\Delta \lambda \over 16\pi G} = {\Delta U \over d}\;,
$$
and we obtain the bubble nucleation amplitude in the limit of weak gravity, with semiclassical  suppression exponent
\eqn\exps{
2W_E\Big |_{{\bar r}\ll1} \approx  v\,B(\sthreehalfs, \sdhalfs)\,\left({d \over \Delta U}\right)^d \,\sigma^{d+1}\;,
}
which agrees with the  known factor of $27\pi^2 \sigma^4 / 2(\Delta U)^3$ in the four-dimensional case.

For large bubbles, where the background curvature effects are felt strongly, we obtain 
\eqn\largeb{
2W_E \Big |_{{\bar r} \gg 1} \approx  q\,v\,B(\sthreehalfs, \sdmtwohalfs)\, {\bar r}^{\,d-2}\;,
}
a form that will be used later in the matching to the dual CFT. 

The same results can be obtained in  an explicitly $O(d+1)$-invariant formalism, more akin to the original presentation of \cdl.  A more detailed discussion of the equivalence of both methods is relegated  to Appendix 1.

\subsec{Decay of excited states}

\noindent

For  excited initial  states, with $M_+ >0$,  we can think of the  branes as thermally emitted by the black hole horizon at $r=r_0$, with a basic rate governed by Hawking's formula, proportional to $\exp(-\beta \omega)$, \foot{At the semiclassical level we are insensitive to the effects of quantum statistics. More generally, we may approximate the Hawking rate by the detailed balance formula $\exp(\Delta S)$, with $\Delta S$ the entropy jump in the emission process. Such a generalization may induce chemical potential terms associated to conserved charges.} with
$\beta(M_+) = 1/T(M_+)$ the inverse Hawking temperature associated to the mass $M_+$. This basic input rate is further convoluted with the potential barriers that the branes may encounter outside the horizon, i.e. we have the `grey-body' spectral formula for the total rate
\eqn\hr{
\Gamma(M_+) \sim \int_{\omega_{0}}^{M_+}  d\omega\, e^{-\beta \omega} e^{-2W_E (\omega)}\;,
}
where $\omega_{0} = -q\,v\,r_0^d$ is the minimum emission energy of branes by the horizon. If the barrier is present, we may approximate this integral by its value at the saddle point to obtain the usual result (cf. \refs\affleck) 
\eqn\aced{
\Gamma(\beta) \propto \exp\left(-I_E (\beta)\right)\;,
}
where $I_E (\beta)$ is the Euclidean action evaluated over periodic Euclidean solutions of \effeu\
with period $\beta (M_+)$. 

For high enough excitation energy $M_+$   the barrier in the potential disappears and all branes with negative
energy $\omega_{0} < \omega < 0$ are emitted without exponential suppression, in which case the formula
\hr\ does not apply. In this situation the decay rate is just like that of a Schwarzschild black hole, i.e. it is given by
the natural time scale of the black hole, $\Gamma(M_+) \sim T(M_+)$. The critical excitation energy beyond which the barrier disappears altogether corresponds to a vanishing sphaleron energy, $\omega_s =0$, and   can be computed by solving $V_{\rm eff} ({\bar r}_c) = V_{\rm eff}' ({\bar r}_c)=0$ for
$\omega=0$, resulting in a critical radius  ${\bar r}_c^2 = (d-2)\alpha^2 / d(1-\alpha^2) = (d-2) {\bar r}^{\,2} /d$. So the barrier is not present for
\eqn\nobarr{
\mu_+ \geq {2 \over d}   \left({d-2 \over d}\right)^{d-2 \over 2} \,{\bar r}^{\,d-2} \;.
}
This results conforms to standard intuition about metastable state decays, namely the barrier would become ineffective if we start with a sufficiently large energy $M_+$ above the metastable `vacuum' (pure AdS$_+$). 

\vskip1cm
\centerline{\epsfxsize=0.45\hsize\epsfbox{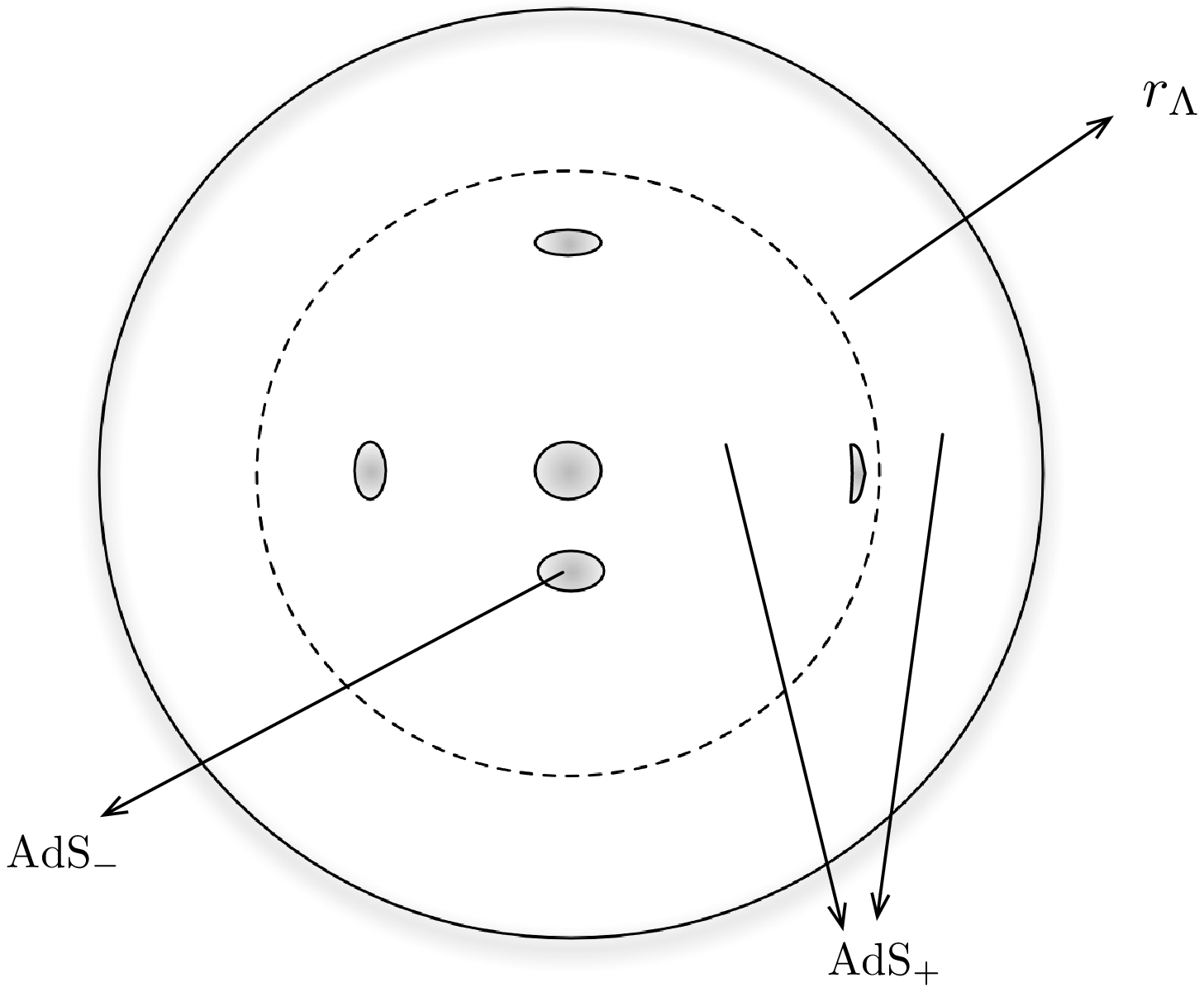}}
\noindent{\ninepoint\sl \baselineskip=8pt {\bf Figure 9:} {\ninerm
Picture of the nucleation process within a sphere of radius $r_\Lambda$. Bubbles of AdS$_-$ grow inside AdS$_+$  and have size ${\bar r} \ll r_\Lambda$ at birth. In the coordinates chosen, only the central bubble looks spherical, although they are all approximately spherical. Most of the bubbles are nucleated in the vicinity of $r=r_\Lambda$ since most of the volume is concentrated there. In QFT terms, these are the `smaller' bubbles, according to the UV/IR dictionary. The bulk volume term implies a UV divergence of the rate. }}
\vskip1cm

\subsec{Measure}

\noindent

Going back to the vacuum transition, with $\omega=0$ and $f(r)=1+r^2$, we may now give an estimate of the preexponential factor. A detailed computation would involve examining the one-loop fluctuation determinant, including its zero (and negative) modes, around the classical Euclidean configuration. We can, however, give an estimate based on the symmetries of the problem.  
Since AdS is a homogeneous space, the rate of nucleation picks a volume degeneracy factor from integration over bubble position inside AdS$_+$ with Euclidean metric 
\eqn\eusc{
ds^2 = (1+ r^2)\,dt_E^2 + {dr^2 \over 1+r^2}  + r^2 \,d\Omega_{d-1}^2
\;.}
Working within a cutoff radius $r_\Lambda$, as illustrated in figure 9,  
 we have a factor of  
$
\Delta t \,m_* \,(m_* \,r_\Lambda)^{d}
$
in the rate, with $m_*$ a microphysical mass scale that arises in the computation of the one-loop functional determinants in the bulk. \foot{This homogeneity property  of the AdS background implies that the leading decay process is still given by vacuum CdL bubbles even in the presence of a black hole, since vacuum bubbles nucleated far from the black hole are approximately described by the zero-temperature effective potential.} In vacuum decay problems, $m_*$  is usually governed by the gap of the metastable state, i.e. $m_* \sim 1/R_+  = 1$.   Hence, we conclude that the decay rate is afflicted by a volume divergence. \foot{The importance of this fact was emphasized to us in early 2009 by Daniel Harlow. See also \polhorow\ for similar remarks in a different context.}

As long as we are interested in the local bulk physics, we can just define the decay rate per unit time and unit volume, obtaining $m_*^{d+1} \,\exp(-2W_E)$.  On the other hand, anticipating the field-theory interpretation through the AdS/CFT dictionary, we may also consider the decay rate per unit time and unit volume of the conformal boundary ${\bf R} \times {\bf S}^{d-1}$,
which diverges as $m_*^{d+1} \,(r_\Lambda)^d$, as always in units $R_+ =1$.  This radial  divergence will be interpreted as a UV divergence in the CFT.

\newsec{Matching to a boundary QFT picture}

\noindent

The description of the shell dynamics in terms of a brane action is appropriate  for the matching to more detailed specifications of the theory, where the bulk theory is actually a string theory, and defects are identified as particular D-branes in the spectrum. The simplest case is the basic blueprint for holography, namely the duality
between type IIB string theory on AdS$_5 \times {\bf S}^5$ with $N$ units of Ramond--Ramond flux and four-dimensional maximally supersymmetric $SU(N)$ Yang--Mills theory. 

A pattern of gauge symmetry breaking of the form $SU(N) \rightarrow SU(N-1) \times U(1)$ is described in the bulk by the dynamics of a probe D3-brane.  
The brane action is identified with the effective action of the $SU(N)$ gauge theory, evaluated over configurations of the adjoint scalar fied $\Phi$ of the form 
\eqn\uuno{
\Phi = {\rm diag} (0, 0, \dots, \phi)\;,
}
 i.e. we have an effective Lagrangian for $\phi$, after we use the UV/IR map $\phi \sim r$, having integrated out all the `unhiggsed'
$SU(N-1)$ degrees of freedom. In the bulk AdS description, crossing a D3-brane amounts to a jump of one unit of Ramond--Ramond flux through the ${\bf S}^5$ factor, which in turn produces a small change in the effective five-dimensional AdS cosmological constant. Hence, we have all the ingredients to interpret spherical D3-brane probes as shells bounding bubbles that mediate transitions between AdS vacua of different curvature radius. We may as well consider a simple generalization in which $n$ D3-branes, with $n\ll N$, are bundled into a bubble wall with $SU(n)$ internal degrees of freedom.

More generally, given an AdS/CFT model constructed from a decoupling limit of $N$ branes, the number of microscopic  degrees of freedom (central charge) 
scales as $N_{\rm eff} \sim N^a$, with $a$ a rational number of $\CO(1)$  (for example, for gauge theories we have $a=2$ and for M2-branes we have $a=3/2$.) It is related to Newton's constant and the curvature radius by $N_{\rm eff} = N^a \sim R_+^{d-2} / G$.  If we imagine that the AdS$_-$ region is obtained from the AdS$_+$ configuration by the removal of  $n$ 
constituent branes (or $n$ units of the corresponding flux in some extra compact manifold) then
we have
\eqn\radiis{
{R_+ - R_- \over R_+ + R_-} \sim {N^{a\over d-1} - (N-n)^{a\over d-1} \over N^{a\over d-1} + (N-n)^{a\over d-1} } \sim {a\over 2(d-1)} {n \over N} 
\;.}
Since $n\ll N$ for the `thin brane' picture to make sense,  this quantity is always small throughout our discussion, quite independently of whether the bubbles are
close to BPS saturation or not, and we shall set $n=1$ and $1\ll N < \infty$ in what follows. In particular, for order of magnitude estimates, we can consider
an average value of the curvature radius $R\sim R_- \sim R_+  =1$.   It is interesting to notice that the large $N$ gauge theory model provides a `mini-landscape', with $\CO(N)$ quasi-degenerate vacua.

Using the relation between $N_{\rm eff}$ and $G$, together with \radiis, we find the scalings 
$$
q \sim N^a \left({1\over  N} - \kappa^2\right) \sim  N^{a-1} -  \alpha^2 q^2
\;,$$ where we have used $R_+ =1$ and $\sigma = \alpha \,q $. From these expressions we can extract the scaling of $q$ and $\sigma$ as a function of  $ N$. In the physical situation  $\alpha^2 /  N \ll 1$ we find 
$
q \sim N^{a-1}
$, 
obtaining back the usual scaling $q \sim 1 / g_s \sim  N$ in the standard case of ${\cal N} =4$, $SU(N)$ SYM theory. 
More detailed comparisons of the effective charge $q$, as defined in \defss, would require taking into account $\CO(1/N)$ jumps of the five-dimensional Newton's constant across a D3-brane in the AdS$_5\times {\bf S}^5$ model, while \defss\ is an effective $(d+1)$-dimensional description, incorporating just vacuum energy jumps.

\subsec{The canonical frame near the boundary}

\noindent

In order to compare the brane motion in the bulk with  standard presentations of CFT degrees of freedom it is convenient to
rewrite the brane effective action in terms of a canonical variable with a standard Lagrangian at low velocities, i.e. we
perform the field redefinition $r(t)\rightarrow \phi(t)$ so that 
\eqn\cano{
I =  v\int dt \left( 
\shalf {\dot \phi}^2 - V_s (\phi) + \CO({\dot \phi}^4) \right)\;,
}
where $V_s (\phi)$ is the `static' potential, defined by $\omega |_{{\dot r}=0} = v \,V_s (r)$. \foot{This is the rigid spherical brane case of more general field redefinitions that may be found in \refs\seibwit.}   Expanding \branea\ to quadratic order in ${\dot r}$ and matching to \cano\ we find the map between the radial variable $r$ and the canonical brane field as 
\eqn\mapcan{
\phi = \sqrt{\sigma} \int_0^r dr'  {(r')^{d-1 \over 2} \over f(r')^{3/4}} = {2\sqrt{\sigma} \over d-2} \;r^{d-2 \over 2} \,\left(1+ \CO(r^{-2})\right)\;,
}
whereas the static potential is given by
\eqn\statp{
V_s (\phi) =  \sigma \,r^{d-1}\, \sqrt{f(r)} - q\,r^d\;,
}
with $r$ solved in terms of $\phi$ by inverting \mapcan. In the large $\phi$ region, it reads
\eqn\largphi{
V_s \big |_{\phi \rightarrow \infty} = {(d-2)^2 \over 8} \,\phi^2 - \lambda\,\phi^{2d \over d-2} + \dots,}
where 
\eqn\defl{
\lambda= \left({d-2 \over 2}\right)^{2d \over d-2}  \,{1\over \sigma^{2\over d-2}} \,{ q-\sigma \over \sigma}\;.}
The first term in \largphi\ gives the conformal coupling 
$$
{d-2 \over 8(d-1)} \,{\cal R}\,\phi^2
$$
 to the background curvature of the  boundary ${\bf S}^{d-1} \times {\bf R}$, and the  second term  is a standard marginal operator in $d$ spacetime dimensions.  Notice that the potential is asymptotically unbounded below whenever
 $\sigma < q$, i.e. precisely when the BPS bound is violated.
 
 The $1/r$ corrections to both the field-theory Lagrangian \defl\ and the field redefinition \mapcan\  are proportional to powers of the background curvature ${\cal R}$, which breaks spontaneously the conformal symmetry. In the limit of a flat boundary, ${\cal R} \rightarrow 0$, both \mapcan\ and \defl\ are given by a single monomial.

We can check this identification by obtaining \mapcan\ directly from the microscopic brane picture.   Let the field $\phi$ be normalized canonically in the D3-brane world-volume action, so that  we have $r/2\pi\alpha' = g_{\rm YM} \phi$ for the mass of a `W-boson' constructed from a stretched string. Here  $\alpha'$ is the type IIB string Regge slope parameter and $g_{\rm YM}^2 = 2\pi g_s$ is the world-volume Yang--Mills coupling in terms of the string coupling constant $g_s$.  Using the fact that the BPS tension of the D3-brane is $\sigma = ((2\pi)^3 g_s \alpha'^{\,2})^{-1}$ we find  perfect agreement with the rule expressed in \mapcan\ for $d=4$. 

The identification \defl\ in the D3-brane theory allows us to trace the effective coupling $\lambda$ back to the SYM Lagrangian. In this case $\lambda  \sim (1-\alpha)/N$ and the operator $-\lambda\, \phi^4$ can be obtained from a standard single-trace
quartic operator $-{1-\alpha \over N} \,\Tr\,\Phi^4 + \dots$, evaluated along \uuno, (we use canonical normalization of the adjoint scalar fields.) On the other hand, we shall obviate details about the R-symmetry structure of the operators or, equivalently, the localization of D3-branes on the ${\bf S}^5$ or analogous `internal Einstein manifolds'.

\subsec{The large-brane CFT and  the Fubini Instanton}

\noindent

Our discussion so far indicates that standard CdL bubbles can be associated to branes, at least to the extent that the thin-wall approximation is applicable to the shells bounding the bubbles. Using the asymptotic map \mapcan\ between the collective coordinate of the shell and  a canonical field in the $d$-dimensional CFT, we find that the CdL dynamics of  (large) spherical shells in AdS$_{d+1}$ is described by nonperturbative bubble nucleation effects of a  conformal $d$-dimensional  Lagrangian of the form 
\eqn\lagft{
{\cal L}_d = -\half (\pt \phi)^2 - {d-2 \over 8(d-1)} \,{\cal R}\,\phi^2 + \lambda \,\phi^{d\over \Delta}\;,
}
where ${\cal R}$ is the Ricci scalar of the $d$-dimensional CFT spacetime and  we denote
$$
\Delta = {d-2 \over 2}
$$
the mass dimension of the scalar field. The effective coupling  of the marginal scalar operator is given by \defl, \foot{It is quite interesting to note  that the scaling of $\lambda$ with $N$ is the same, equal to $1/N$, in all three `canonical' constructions of CFTs from parallel branes, namely D3, M2 and M5 branes.}
 \eqn\effcc{
 \lambda =  \Delta^{d/\Delta} \,\sigma^{-{1\over \Delta}} \; {q-\sigma \over \sigma}  \sim \left({1 \over N^{a-1}}\right)^{1\over \Delta} \, {1-\alpha \over \alpha^{d \over 2\Delta}}\;.
 }
  Hence, the violation of the BPS bound for the bulk branes, $\alpha < 1$, is equivalent to  $\lambda >0$, namely the condition of {\it instability} of the CFT potential at large values of the field. Notice that this detailed map with the particular interactions in \lagft\ and \effcc\ is actually 
  asymptotic in the sense that it was derived for bubbles of large size, ${\bar r} \gg 1$, in units of
  the AdS$_+$ radius (cf. \mapcan). In principle, one may extend the analysis to general ${\bar r}$ in units of $R_+$, at the expense of keeping all terms in the power expansions of \mapcan\ and \statp. 
  
Classical solutions of \lagft\ with different values of the constant background curvature, ${\cal R}$, can be related by conformal transformations. In particular, the explicit map
\eqn\confmap{
dt_E^2 + d\Omega_{d-1}^2 = {1\over u^2} \left(du^2 + u^2 \,d\Omega_{d-1}^2\right)
\;,
}
with $|x| = u = \exp(t_E)$, expresses the conformal equivalence of the Hamiltonian manifold ${\bf R}\times {\bf S}^{d-1}$ with the flat hyperplane 
${\bf R}^d$. Using this map we can profit from the knowledge of instanton solutions of the massless theory, defined on ${\bf R}^d$. Such instantons are Euclidean versions of field 
  configurations discussed by Fubini in \refs\fubini\  (see also \refs\zinn\ for a review) 
and take the form 
\eqn\inst{\phi_{\rm inst} (x) = \sqrt{2\over \lambda} {\rho \over |x|^2 + \rho^2}\;,
}
in the four-dimensional case, where $|x|$ is the Euclidean length on ${\bf R}^4$ and $\rho$ is an arbitrary length scale that characterizes the `size' of the instanton, with action
$S_{\rm inst} = 2\pi^2 / 3\lambda$. The similarity to Yang--Mills instantons is clear from the explicit formula \inst\ and, like in the gauge theory case, they {\it do not} have thin walls, despite being characterized by a size parameter. \foot{We shall refer to these Euclidean configurations as `Fubini instantons', despite the fact that they are actually bounces (i.e. having one negative eigen-mode). In particular, these solutions are used as approximations to   false vacuum bounces  in situations where the thin-wall approximation is not applicable, cf. \refs\linde.} These instantons mediate the decay of the classical $\phi=0$ state by
nucleation at $t=0$ of bubbles
$$
\phi_{\rm bubble} ({\vec x}, t=0) = \sqrt{2\over \lambda} {\rho \over {\vec x}^{\;2} + \rho^2}
$$
 of size $\rho$ and field value $ { \phi_0} \sim 1/\sqrt{\lambda \rho^2}$ at the center, that subsequently expand in an asymptotically null trajectory. Notice that the energy barrier inducing the metastability of the $\phi=0$ configuration is supported  just by kinetic terms in the massless model.

All these considerations can be generalized to arbitrary  $d>2$ dimensions. The instanton solution on
${\bf R}^d$ with size $\rho$ and position $x_0$ parameters is now given by 
\eqn\linstd{
\phi_{\rm inst} (x)= \left({2\over \lambda}\right)^{\Delta / 2} \left({\Delta \,\rho \over |x-x_0 |^2 + \rho^2}\right)^{\Delta}\;,
}
and the resulting instanton action
\eqn\insac{
S_{\rm inst} = \int_{{\bf R}^d} \left(\shalf(\pt \phi_{\rm inst})^2 -\lambda\,(\phi_{\rm inst})^{d / \Delta}\right) = {  2^{d/2} \Delta^d \,v \over\lambda^\Delta} \int_0^\infty ds\,{s^{d-1} (s^2 -1) \over (s^2 +1)^d}\;,
}
where, as before, $v= |{\bf S}^{d-1}|$. 

We can evaluate the integral in terms of special functions as follows. We first split the $s^2 -1$ term in the numerator of \insac\ and work out each integral separately with the change of variables
$s \rightarrow (1+s^2)^{-1}$, resulting in integral representations of beta functions, to finally  find
\eqn\ccal{
S_{\rm inst} = { c\over \lambda^{\Delta}}\;, \qquad {\rm with}\;\;\;\;
c= {v \over 2^{\Delta}} \Delta^d \, B(\sthreehalfs, \sdmtwohalfs)\;.
}
Alternatively, the same action can be computed by using a stereographic projection to  map the ${\bf R}^d$  problem into the same problem  on ${\bf S}^d$, with
curvature scalar ${\cal R} = d(d-1)$. In this case, a constant solution of the Euclidean equations of motion is
${\bar \phi} = (\Delta^2 /2\lambda)^{\Delta /2}$, with action
\eqn\acscuatro{
S_{\rm inst} = |{\bf S}^d| \left( {d(d-2)  \over 8} {\bar \phi}^{\,2} - \lambda \,{\bar \phi}^{\,{2d \over d-2}} \right)= {c \over \lambda^\Delta}\;,
}
where we have used the explicit formula for the spheres' volume, $|{\bf S}^{d-1}| = 2\pi^{d/2} /\Gamma(d/2)$.

In order to compare these instantons to our CdL bubbles in the bulk, we can apply the conformal transformation  \confmap\ to \linstd\ and obtain the corresponding instanton fields on ${\bf R}\times {\bf S}^{d-1}$. For general values of $x_0$ one gets a complicated expression, representing  an instanton field partially localized on the sphere. A simpler configuration is obtained for $x_0 =0$, with the form  
\eqn\inssphere{
{\tilde \phi}_{\rm inst} (t_E, \Omega) = u^\Delta \,\phi_{\rm inst} (u) = \left({\Delta^2 \over 2\lambda}\right)^{\Delta/2} \left(\cosh((t_E - t_\rho)\right)^{-\Delta}\;,
}
where $t_\rho = \log \rho$ acquires the interpretation of Euclidean time location rather than size, since the resulting instanton on ${\bf R} \times {\bf S}^{d-1}$ is constant on the sphere with value ${\bar \phi} = (\Delta^2 /2\lambda)^{\Delta /2}$ at the time-symmetric point. 

\subsec{A detailed bulk/boundary matching}

\noindent

We check the duality of Fubini instantons with bulk CdL bounces in three stages. First, we notice that the  field value  of the instanton, ${\bar \phi}$, exactly matches the nucleation radius of the CdL bubble, ${\bar r}$, according to the map
\mapcan\ in the $\alpha \rightarrow 1^-$ limit. This follows by direct inspection of \mapcan, using the formula \effcc\ in the mentioned limit. Second, we are able to exactly match the instanton action \ccal\ to the previously computed tunneling rate exponent $2W_E$, again  in the $\alpha\rightarrow 1^-$ limit in \largeb. 

To see this, use 
$$
q\,{\bar r}^{\,d-2} = q \left({\alpha^2 \over 1-\alpha^2}\right)^{\Delta} \longrightarrow  {\Delta^d \over (2\lambda)^{\Delta}}
$$
to finally obtain
\eqn\bulkw{
2W_E \longrightarrow q \,v\,{\bar r}^{\;d-2} \,B(\sthreehalfs, \sdmtwohalfs) \longrightarrow {c \over \lambda^{\Delta}}\;,
}
with the same constant, $c$, previously  evaluated in \ccal. 
The precise matching of the leading exponential rate of nucleation is quite remarkable, since
no supersymmetry can be summoned to explain the precise agreement. The required bubbles {\it must} violate the BPS bound in order to have a nonzero nucleation rate. In turn, this means that
the vacua in question have to break supersymmetry.

The third check involves the measure over the instanton moduli space.  In the single-instanton sector, conformal invariance fixes the instanton measure completely implying a rate proportional to (in ${\bf R}^d$ variables) 
\eqn\meass{
 d^d x_0  \,{d\rho \over  \rho^{d+1}} \exp\left(-c/\lambda^{\Delta}\right)
\;.}
Postponing for the time being the question of convergence of \meass, we notice that the same expression can be reproduced in the bulk description.  Any two instanton configurations in the CFT can be obtained from one another by a conformal transformation (such as translations and dilatations in the form of \linstd). Therefore, having successfully matched particular instanton configurations which are uniform on ${\bf S}^{d-1}$ or ${\bf S}^d$, we are guaranteed a complete matching throughout the whole moduli space, since the conformal group is realized as an isometry group of AdS in the bulk description. 

We have presented the discussion of spherical bubbles in section 2 in a particular coordinate system with explicit $U(1)\times O(d)$ isometries, adapted to the boundary CFT geometry ${\bf R} \times {\bf S}^{d-1}$ with the first factor representing the time direction. Hence,   bubbles centered at $r=0$ represent homogeneous configurations on the CFT spatial sphere. However, the
bulk AdS spacetime is completely homogeneous, so that bubbles of the same proper size (controlled by ${\bar r}$) will nucleate homogeneously throughout AdS$_{d+1}$ with uniform probability per unit  AdS volume.  A bubble  nucleating `off center' with respect to the $O(d)$-symmetric frame, say centered around the point with coordinates $(r_b, \Omega_b)$,  represents an inhomogeneous configuration on the CFTs'
${\bf S}^{d-1}$ spatial sphere.

 We can actually exhibit the form \meass\ of the measure by going to a boundary of the instanton moduli space corresponding to $r_b \gg 1$. In this limit, the standard UV/IR relation of AdS/CFT applies,  and such CdL bounce will be interpreted as   a Fubini instanton centered around  $(t_b, \Omega_b)$ and with a `boundary size'  given by  $\rho \sim 1/r_b$ (cf. figure 10). 
This relation implies that bulk and boundary single-instanton measures are actually equal
\eqn\mattt{
 dt_b \,
d\Omega_b  \,dr_b\, r_b^{d-1} \; \exp\left(-2W_E \right) \approx d^d x_0 \, {d\rho \over \rho^{d+1}} \,\exp\left(-c/\lambda^\Delta\right) \;,
}
in the $r_b \sim \rho^{-1} \gg 1$ region of the moduli space,  since small instantons do not distinguish ${\bf R}\times {\bf S}^{d-1}$ from ${\bf R}^d$. It would be interesting to extend the quantitative bulk/boundary checks to the value of the negative eigenvalue for quadratic fluctuations around the bounce solutions.

\bigskip
\centerline{\epsfxsize=0.45\hsize\epsfbox{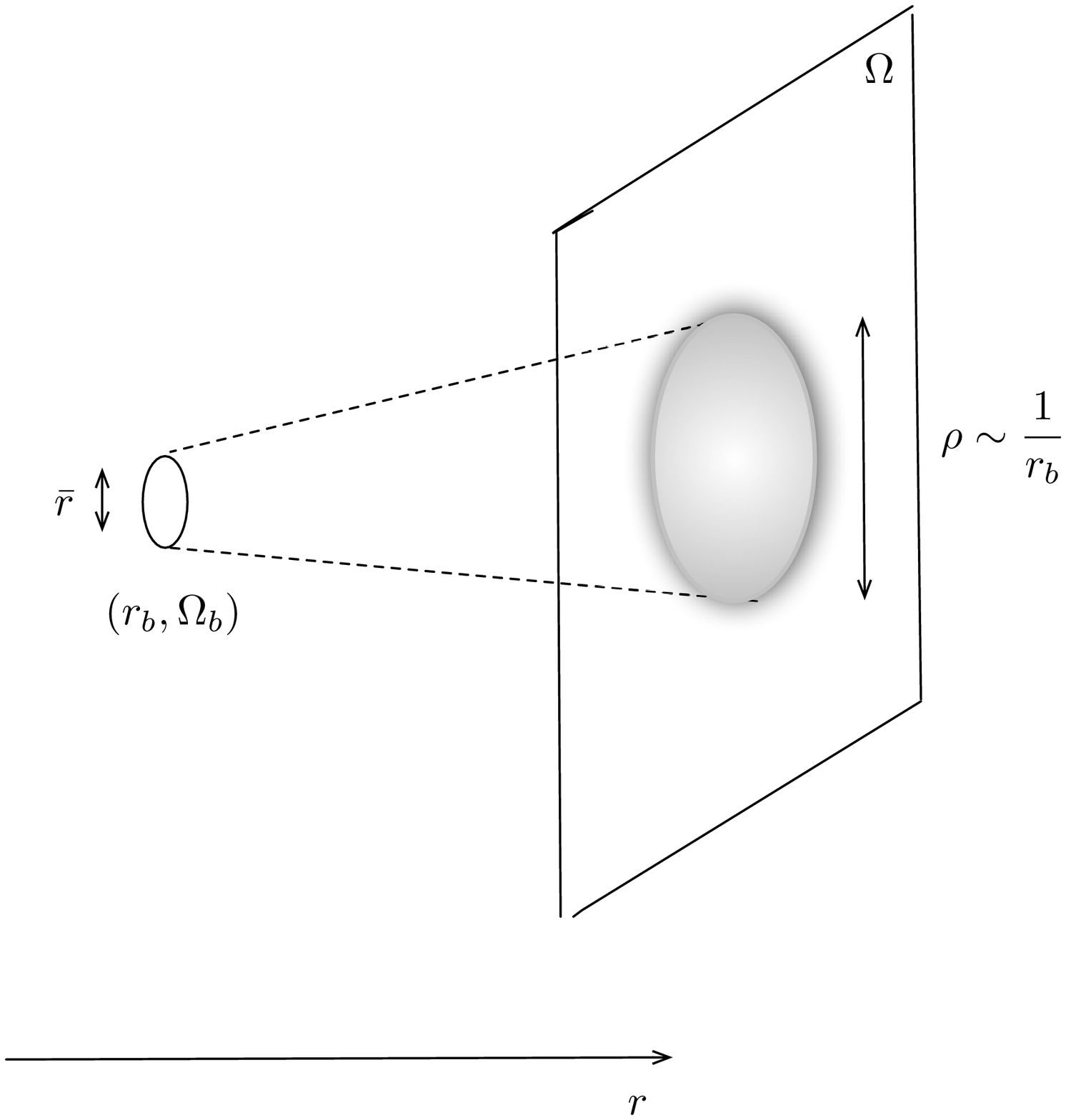}}
\noindent{\ninepoint\sl \baselineskip=8pt {\bf Figure 10:} {\ninerm
A bulk CdL bubble of size ${\bar r}$ and sitting at the point $(r_b, \Omega_b)$, with $r_b \gg 1$, is dual to a Fubini bubble of size $\rho \sim 1/r_b$ on the boundary theory, centered around $\Omega_b$. }}
\bigskip

The  singularity of the   measure \mattt\  at $\rho =0$ leads to a divergent rate in the dilute-instanton approximation, a characteristic feature of CFTs.  In the bulk description, the singularity has a simple interpretation from integrating the nucleation point of the bubble over the infinite volume of AdS. This singularity may be cured by fermion zero modes, forcing potentially divergent amplitudes to vanish,  a mechanism at work  in supersymmetric CFTs.  If the UV fixed point is replaced by an asymptotically free theory the UV divergence may be turned into an IR  `large instanton' divergence, as happens for example in Yang--Mills theories.  A safer possibility is to  contemplate a modification of the UV behavior that effectively cuts off the zero-size region of instanton moduli space. For instantons mediating vacuum decay, the natural UV modification involves then a {\it stabilization} of the theory. 

We shall see in the next section that the divergence of the decay rate is related to the crunch singularity that ensues when considering the classical evolution after nucleation.

\newsec{ Remarks on crunches and their regularization}
\noindent

Having identified a fairly precise duality between CdL bounces and Fubini instantons in an unstable CFT, we may now consider the fate of the nucleated bubbles as they grow towards the boundary of AdS. As explained above, these zero-energy  bubbles reach the boundary of AdS in finite asymptotic time, and $dr/dt \sim d\phi /dt$ diverges in the process. This is interpreted in the dual CFT as the roll down in a  potential of an unstable marginal operator. 

 The precise value  of the time elapsed between nucleation and the arrival at the  boundary does depend on the nucleation point $r_b$. The UV/IR correspondence translates this $r_b$ dependence into the size parameter of the CFT bubble, $\rho \sim 1/r_b$,  and to  the central value of the field,   via the relation $\phi_0 \sim 1/\sqrt{\lambda \rho^2}$. This makes contact with the results of  \refs\rabroll, where  it was found that the  time to roll down a conformally invariant potential was only a function of the starting point $\phi_0$.  Working on ${\bf S}^3 \times {\bf R}$, the minimal value of  $\phi_0$ is ${\bar \phi} = (\Delta^2 /2\lambda)^{\Delta/2}$,  which translates into ${\bar r}$ with the use of \mapcan\ giving  the maximal hit time computed in \hit, corresponding to a `centered' bubble nucleation.
 
 The diverging kinetic energy of the rolling scalar field triggers an UV singularity  which betrays the presence of a crunch singularity in the bulk. 
In fact, the structure of this crunch singularity is quite intrincate. The volume extensivity of the decay rate implies that an expanding bubble is bound to collide with an infinite number of other bubbles on its way to the boundary, and moreover
do so in a {\it finite} time. Each collision will release energy in the form of radiation and make the approach to the crunch a rather complicated foam-like process.  In the dual CFT picture we see that smaller and smaller Fubini bubbles are created at a conformally invariant rate, bubbles within bubbles with a fractal-like structure of field strenghts.  \foot{For a recent numerical evaluation of a similar process, in the context of the so-called `spinodal' transition, see for example \refs\marga.} 

The fact that the QFT Hamiltonian following from \lagft\ is unbounded below means that this complicated  decay process  has no identifiable `endpoint' in a nonperturbative sense, i.e. in the QFT space of states, and thus there is no clear suggestion as to how the `crunch' should be treated in the boundary QFT. \foot{However, somewhat formal proposals have been advanced in \refs\turok.}

A natural definition from the physical point of view is to modify the UV behavior of \lagft\ in such a way that the theory is ultimately stable, albeit with a large tunable hierarchy between the stabilization mechanism and the instability described by
\lagft. In such a set up we can give a description of the endpoint and see what became of the crunch. Equivalently, we can
see how the regularized model develops a  `crunching' behavior as the UV regularization is removed. Previous work along these lines includes \refs{\hcc}. Here, we shall address this question emphasizing the requirement of having a well defined exact CFT description of the far UV behavior of the model. 

\subsec{A quantum quench}
\noindent

Let us denote by $\Phi$ the collection of fields in the  QFT, including the brane collective coordinate $\phi$.  Consider  a QFT potential ${\cal V}(\Phi)$  with the property that, when restricted to the $\phi$ direction in field space, it shows the  qualitative features  of the $V(\phi)$ potential depicted in figure 11. This potential supports tunneling phenomena, where the field value of the maximal-size bubbles is ${\bar \phi}$,  and we assume the slope at $\phi > {\bar \phi}$ to be approximately conformal, so that the decay of the semiclassical state at $\phi=0$ proceeds by nucleation of Fubini-type bubbles. Conformal symmetry is broken by $\CO(1)$ effects at  the scale $\phi_w \gg {\bar \phi}$, where the potential is  stabilized with a net drop of potential energy $\Delta V = V(0) - V(\phi_w) = V({\bar\phi})-V(\phi_w)$. Thus, we assume that the region $0<\phi \ll \phi_w$ is described by \lagft\ and controlled by conformal invariance. We also assume that at $\phi \gg \phi_w$ the potential ultimately  retains the conformal properties, being  dominated by a stable marginal operator i.e.  $V(\phi\rightarrow \infty) \rightarrow + \phi^{d/\Delta}$. 

\vskip1cm
\centerline{\epsfxsize=0.46\hsize\epsfbox{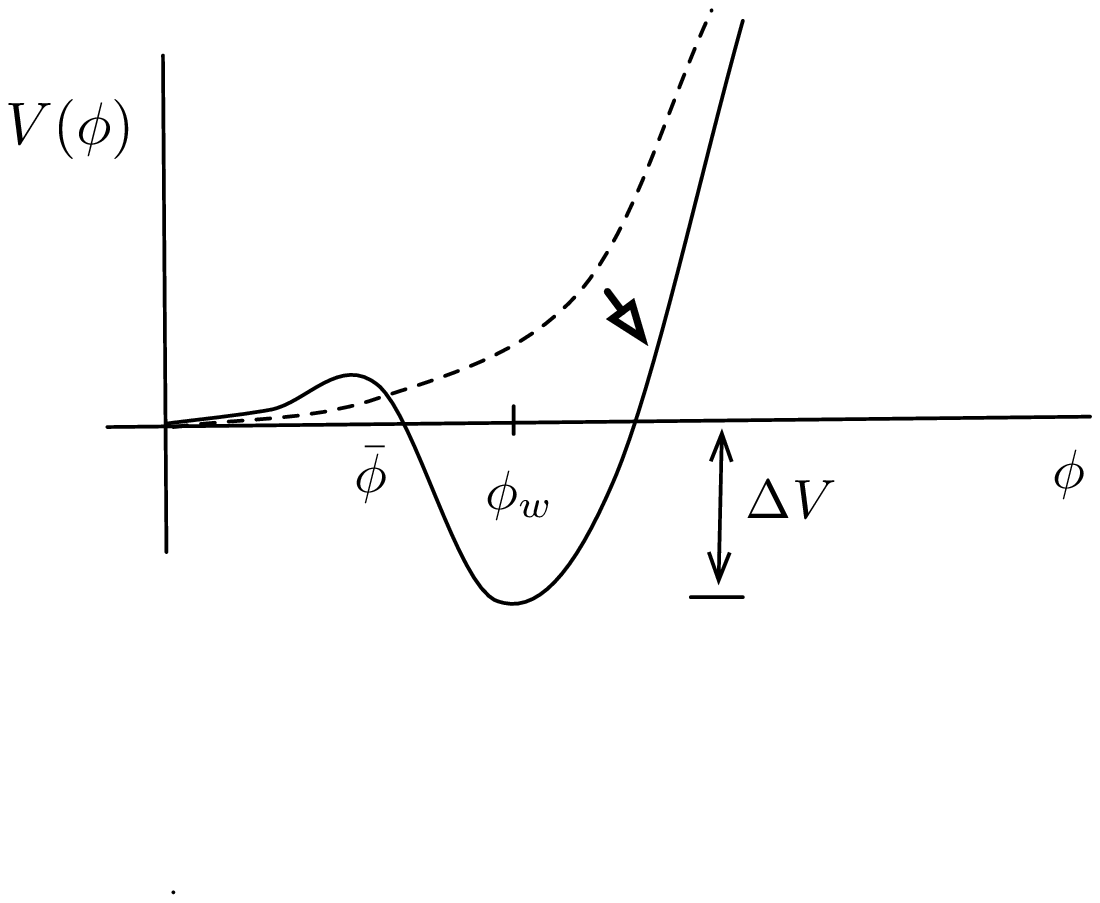}}
\noindent{\ninepoint\sl \baselineskip=8pt {\bf Figure 11:} {\ninerm
QFT potential with tunneling phenomena. Maximal size bubbles are nucleated with typical energy scale ${\bar \phi}$ and the stabilization scale at $\phi_w$ may be set at a hierarchically larger energy. The metastable state dual to AdS$_+$ may be prepared at $t=0$ as the vacuum of the  stable conformal potential in dashed lines. The evolution of this state past $t=0$ (the quantum quench indicated by an arrow) will proceed by nonperturbative bubble nucleation. These bubbles will be of Fubini type provided the unstable section of potential, for $0< \phi < \phi_w$,  is approximately conformal and well described by \lagft.}}
\vskip1cm

Rather than describing the metastable state by the $\phi=0$ configuration, we can think more broadly in terms of a  wave functional $\Psi_+ [\Phi]$ for all CFT degrees of freedom, whose restriction to the  $\phi$ direction in field space is assumed to be  well peaked around $\phi=0$. The basic hypothesis of the AdS/CFT correspondence is that this state admits a semiclassical description in the large $N$ limit in terms of weakly coupled gravity on a background $X_+$ whose geometrical symmetries encode the quantum symmetries of $\Psi_+[\Phi]$. In particular, if $\Psi_+ [\Phi]$ is conformally invariant, $X_+$ is isometric to AdS, and we denote it by AdS$_+$.

We can specify the $X_+$ state more graphically by means of  a `quantum quench'. Let us consider a CFT with an exactly marginal potential ${\cal V}_+ (\Phi)$ whose projection in the $\phi$ direction is the operator $+\phi^{d/\Delta}$. The ground state of this theory is a conformally invariant state that we denote $\Psi_+ [\Phi]$, with a dual description given by the AdS$_+$ background. The quantum quench consists of the sudden change of the potential ${\cal V}_+ (\Phi) \rightarrow {\cal V} (\Phi)$ at time $t=0$, as shown in figure 11. After the quantum quench, the AdS$_+$ state is no longer stationary in the deformed CFT with potential ${\cal V}(\Phi)$, and will decay, in this case nonperturbatively via bubble nucleation.  

An elementary event of bubble nucleation is described by the discrete jump $X_+ \rightarrow X_*$, where $X_*$ is a new background with a bubble, which subsequently evolves perturbatively. The long-time limit of $X_*$ should be generically described as the dual of a locally thermalized state with energy $\Delta V$, measured with reference to the absolute ground state $\Psi_w [\Phi]$ of the QFT defined by ${\cal V}(\Phi)$, after the quantum quench. The detailed description of the decay of $\Psi_+ [\Phi]$ is necessarily very complex, as it involves multiple bubble nucleation and subsequent dissipation and collision of bubbles, but the basic elements of the process can be understood in the terms just described. 

The quantum quench construction is just a formal device to identify initial states with particular properties, in this case the conformal nature of the initial AdS$_+$ state. For our purposes, the important property of this procedure is  the bounded nature of the  potential deformation ${\cal V}_+ \rightarrow {\cal V}$, which has  
 `compact support' in field space, namely the two potentials have the same UV asymptotics. This ensures that the ground state of the new potential,  denoted $\Psi_w [\Phi]$, has the same UV asymptotic behavior as $\Psi_+ [\Phi]$, and moreover the two states differ only by a finite amount of energy. On general grounds, all normalizable states of the CFT with finite energy will share this property, namely they all look like the vacuum when probed at very short distances. As an immediate corollary of this statement, the bulk background $X_w$, dual to the new ground state $\Psi_w [\Phi]$, will have the same asymptotics as $X_+$, namely it must approach AdS$_+$ near the boundary. 

This simple argument,  using basic facts about quantum field theory and the AdS/CFT correspondence, shows that  no finite-energy state  can ever change its boundary asymptotics as a result of a decay process. By {\it finite energy} we of course mean the energy measured with respect to the true ground state of the theory.\foot{This true ground state, $X_w$, has negative ADM energy when measured with respect to that of AdS$_+$, thus violating the corresponding  positive energy theorem   \abde.  This is possible by evading the required energy condition, since $U(\chi)$ can attain  negative values below  $U_+$ in the bulk gravity Lagrangian, precisely for those configurations with (non-supersymmetric) AdS$_-$ bubbles.} In particular, the AdS$_+$ $\rightarrow $ AdS$_-$ transitions suggested by the evolution of CdL bubbles are only possible to the extent that they represent {\it infinite} energy falls, as in the unstable model \lagft. Thus, this embedding of the problem into a well-defined AdS/CFT model puts into perspective the occurrence of crunches in the final state. 

\vskip1cm
\centerline{\epsfxsize=0.5\hsize\epsfbox{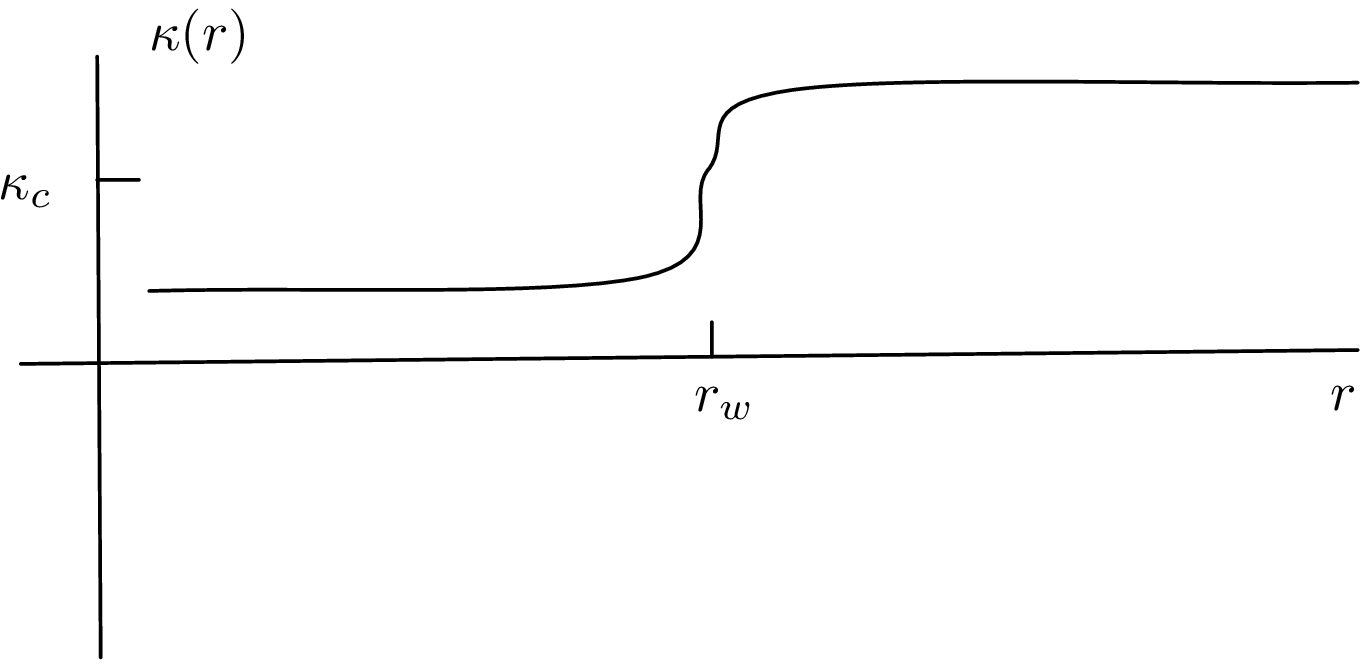}}
\noindent{\ninepoint\sl \baselineskip=8pt {\bf Figure 12:} {\ninerm
Running value of the bubble tension. We assume that it jumps from the under-saturated regime $\kappa < \kappa_c$ at $r<r_w$ to the over-saturated regime $\kappa > \kappa_c$ at $r>r_w$. The threshold radius $r_w$ is a free parameter in the bulk picture. For the sake of clarity, we choose
$r_w$ much larger than any other length scale in the bulk description.}}
\vskip1cm

In the following sections we describe a qualitative scenario for a CFT stabilization and its effects on the crunching problem, using the bulk effective brane description. Then in section 5.4 we discuss  some possible routes towards the detailed construction of the potential $V(\phi)$ of figure 11 in concrete AdS/CFT examples.

\subsec{A bulky stabilization}
\noindent

The general remarks in the previous section indicate that any successful stabilization of the CFT will incorporate an energy scale $\phi_w$, with a bulk radial scale counterpart $r_w \sim (\phi_w /\sqrt{\sigma}\,)^{2\over d-2}$, beyond which no bubble nucleation takes place and moreover bubbles nucleated at lower scales are stopped in their rolling towards the UV. \foot{This strategy of making the asymptotic geometry `safe' was used to good effect in other problems of AdS stabilization (cf. for example  \bekus).}

We can devise a simple bulk model that incorporates these features by assuming that $r_w$ is the location of a bulk domain wall for the effective brane tension.  Namely we have $\sigma < q$ for $r< r_w$ and $\sigma > q$ for $r>r_w$. Equivalently, we have $\kappa < \kappa_c$ for $ r< r_w$ and
$\kappa > \kappa_c$ for $r>r_w$, as shown in figure 12. 

 \vskip1cm
\centerline{\epsfxsize=0.4\hsize\epsfbox{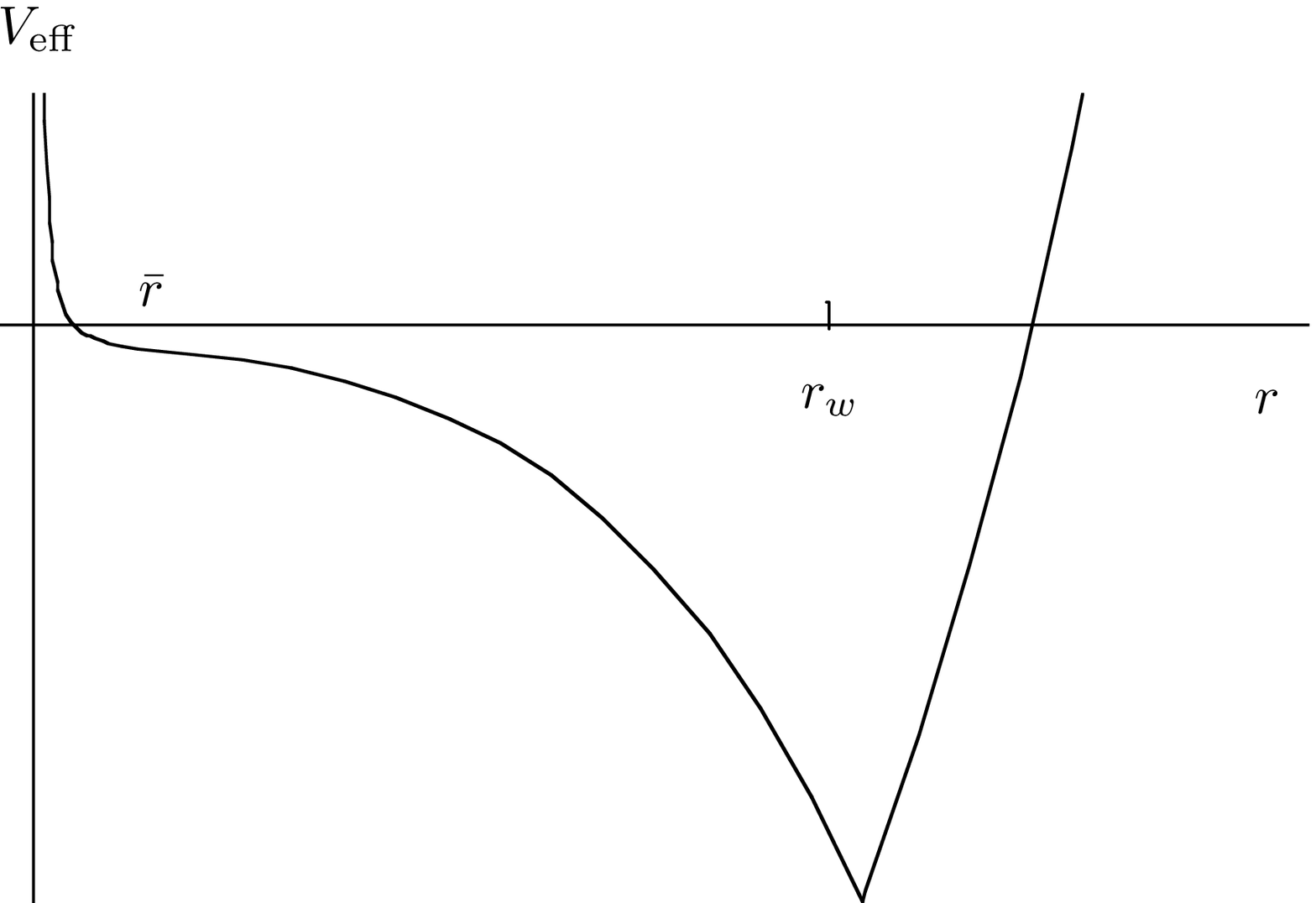}}
\noindent{\ninepoint\sl \baselineskip=8pt {\bf Figure 13:} {\ninerm
Effective potential for a CdL bubble in a regularized theory with stabilization scale $r_w \gg {\bar r}$. A bubble nucleated with radius  at the turning point ${\bar r}$  will enter an oscillatory motion around $r_w$, which eventually will be damped by radiation emission and collision with other bubbles.}}
\vskip1cm

The effective brane potential for this situation can be obtained  by simply patching the low and high tension potentials across the domain wall location, namely we define  $V_{\rm eff} (r) = V_{\rm eff} (r)_{\sigma < |q|}$ for
$r< r_w$, and $V_{\rm eff} (r) = V_{\rm eff} (r)_{\sigma > |q|} + \Delta_w$ for $r>r_w$, with the constant $\Delta_w$ defined by
$
\Delta_w= V_{\rm eff} (r_w)_{\sigma < |q|} - V_{\rm eff} (r_w)_{\sigma> |q|} 
$, 
to ensure that $V_{\rm eff}$ is continuous at $r_w$. A sketch of such a potential is  offered in figure 13.

We may ask if  the `tension domain wall' envisaged here  admits a simple five-dimensional supergravity incarnation, a sort of `toy landscape' description  in terms of an effective potential. A minimal qualitative model would involve a potential
for two scalar fields $U(\chi, \sigma)$, with $\chi$ controlling the value of the cosmological constant and $\sigma$
a dilaton-like field determining the tension of branes. A potential with four local minima, with the form of Fig. 14,   would do the job
 provided it supports domain walls, located at $r=r_w$, separating two regions with tensions $\sigma_\pm$, where $\sigma_- < q$ and
$ \sigma_+ >q$. In addition, the transitions $\chi_+ \rightarrow \chi_-$ should correspond to dynamical bubbles if occurring on a region $\sigma = \sigma_-$ and to static domain walls for the case $\sigma = \sigma_+$.  This can be achieved if the vacuum energy differences satisfy $U(\chi_+, \sigma_-) - U(\chi_-, \sigma_-) > U(\chi_+, \sigma_+) - U(\chi_-, \sigma_+)$, so that the corresponding shells violate the BPS bound in the $\sigma_-$ region but 
 satisfy it in the large tension region $\sigma_+$. 
 
  \bigskip
 \centerline{\epsfxsize=0.5\hsize\epsfbox{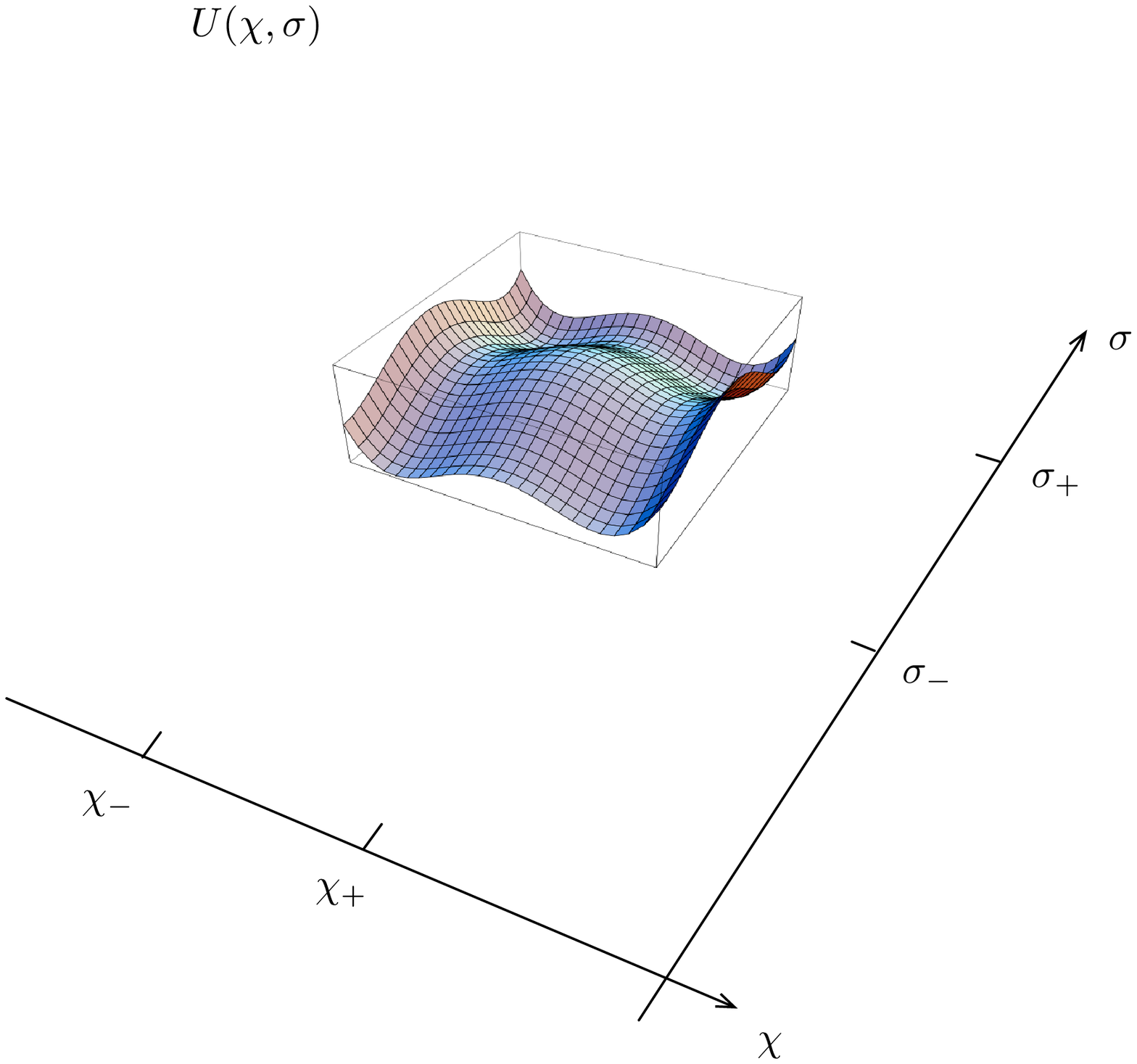}}
\noindent{\ninepoint\sl \baselineskip=8pt {\bf Figure 14:} {\ninerm
Schematic potential supporting a domain wall for the value of the shell tensions and vacuum bubbles. CdL bubbles correspond to transitions $(\chi_+, \sigma_-) \rightarrow (\chi_-, \sigma_-)$. On the other hand, the vacua $(\chi_+, \sigma_+)$ and $(\chi_-, \sigma_+)$ are approximately degenerate, so that no expanding bubbles can exist in a region with $\sigma=\sigma_+$. The regularized background in the text implies considering a combination of CdL bubbles and a static domain wall effecting the transition $\sigma_- \rightarrow \sigma_+$ at $r=r_w$. }}
\bigskip

 In this model, the homogeneous vacua $(\chi_\pm, \sigma_+)$ are stable, protected by the high tension of the branes they support. The lower homogeneous vacuum, $(\chi_-, \sigma_-)$ is also stable because it does not have other vacua to decay into. On the other hand the homogeneous AdS$_+$ background  $(\chi_+, \sigma_-)$ is unstable towards nucleation of bubbles of the $(\chi_-, \sigma_-)$ vacuum, giving the standard CdL decay process ending in a crunch.  The regularized background with bubble instabilities corresponds to a domain-wall spacetime
 which looks like the vacuum $(\chi_+ , \sigma_-)$ for $r<r_w$, and like the vacuum $(\chi_+ , \sigma_+)$ for
 $r>r_w$. Hence, bubbles cannot form in the $r>r_w$ region, but the decay of the $(\chi_+, \sigma_-)$ region of the vacuum proceeds as before. Being a compact domain of the original spacetime, we do not expect any  naked singularities to emerge in this case.

\subsec{Endpoints}

\noindent

According to the effective potential in figure 13, bubbles nucleated at ${\bar r} \ll r_w$ will start oscillating  around $r_w$ and eventually lose their energy into radiation, by settling down to the absolute minimum at $r=r_w$. 
 As the oscillating bubble wall loses energy to radiation, gradually the mass parameter of the interior solution, $ \mu_-$, begins to increase. Hence, the dissipation of oscillation energy into radiation leads to $\Delta \mu <0$ and accordingly  to $\omega <0$ for the slowing-down brane. The effect on $V_{\rm eff}$ is to gradually lift the minimum at $r=r_w$ and bring, at the same time, the pole of the potential into the physical region of positive radii, as illustrated in figure 15.  As the turning point ${\bar r}$ increases, the static brane with no oscillation energy left corresponds to the situation where
$V_{\rm eff} (r_w)=0$, which happens at a negative critical value of the energy  $-\omega_r$, given by
\eqn\zeen{
-\omega_r  = V_s (r_w) = -q\,v\,(r_w)^d + \sigma\,v\,(r_w)^{d-1} \,\sqrt{f(r_w)}\;,
}
where $V_s (r)$ is the static potential defined in \statp. Since we assume $r_w \gg 1$, this `reheating' energy, released in the damping of the brane oscillations, is of order $\omega_r \sim (q-\sigma)\,v \,(r_w)^d = q\,v\,(1-\alpha)\,(r_w)^d \sim \eta \,N_{\rm eff} \, (r_w)^d$, with $\eta = (1-\alpha)/{ N}$. We can interpret this energy as a fraction $1/{N}$ of the plasma energy at effective temperature $T_w \sim r_w$, with a further suppression by the saturation factor $1-\alpha$. 

More generally, using the map between the brane static potential and the CFT potential in the saturation limit, $1-\alpha \ll 1$, we obtain 
\eqn\fallp{
\omega_r \sim \lambda\,(\phi_w)^{2d \over d-2} \sim |V(\phi_w)| = \Delta V\;,}
showing that  $\omega_r$ is also the `reheating' energy in QFT terms.

The bubble nucleation rate scales again as $(r_\Lambda)^d$, with the cutoff replaced by
the stabilization scale, $r_\Lambda \sim \Lambda \sim (\phi_w /\sqrt{\sigma}\,)^{2\over d-2}$, which breaks conformal invariance and renders the instanton measure well defined. In particular, notice that $\sigma > q$ for $r>r_w$, which precludes the nucleation of any brane in this region.

The picture presented so far, where the false vacuum  energy is released 
by dissipation of a coherent oscillation of a single bubble, is necessarily simplistic, since we know that the dominant process of AdS$_+$ to AdS$_-$ conversion involves copious bubble collisions, particularly for the case ${\bar r} \ll r_w$. This means that most of the bubbles will release part of their kinetic energy in collision-induced radiation while they are still falling, far from the minimum at $r_w$.  In other words, each individual bubble will contribute less to the reheating, but there will be many bubbles colliding and  eventually contributing to  a locally thermalized state with a reheating energy $\omega_r \sim |V(\phi_w)|$. 

\vskip1cm
\centerline{\epsfxsize=0.5\hsize\epsfbox{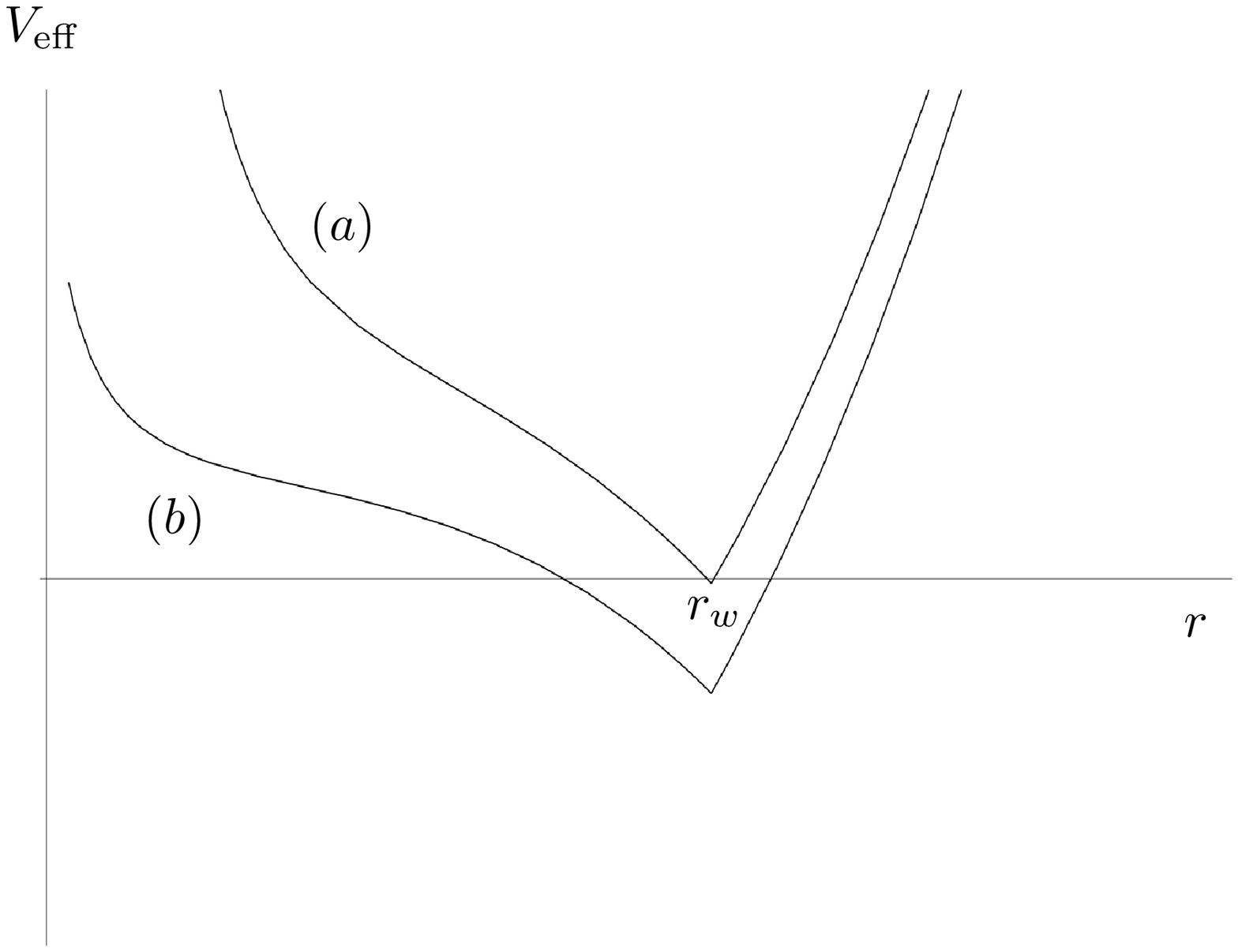}}
\noindent{\ninepoint\sl \baselineskip=8pt {\bf Figure 15:} {\ninerm
Evolution of the effective potential for the bubble wall stabilized at $r_w$,  as the shell loses energy to radiation. The curve $(a)$ corresponds to the brane at rest,  with $M_+ =0$ and $\omega= -\omega_r <0$. It is the result of 
the oscillating shell having shed all its kinetic energy into radiation, thereby stopping the oscillation. The lower curve $(b)$ shows an intermediate case with a partial energy loss.}}
\vskip1cm

The final form of the bulk endpoint configuration depends on the value of the parameters. 
For $\eta\, r_w^d  \gg 1$ the reheating energy is above the threshold for the large AdS black hole, so the radiation
will collapse into a large AdS black hole with radius 
$
r_s \sim  \eta^{1/d} \,r_w  \ll r_w
$. On the other hand, if $\eta$ is so small that $\eta\, r_w^d \ll 1$ the interior black hole will be
small in units of the AdS curvature radius and we see that in either case the resulting black hole is well-contained inside the sphere of radius $r_w$. 
  In fact, when $\eta$ is too small,
it is entropically favorable for the system to settle into a graviton radiation gas. This happens
for released energies below the threshold
$
\omega_{ r} \ll (N_{\rm eff})^{d+1 \over 2d-1} 
$, 
again in units of $R_+ =1$. 
The corresponding bound on $\eta$ for the pure radiation endpoint is
$
\eta \ll (N_{\rm eff})^{2-d \over 2d-1}
$. 

For a fixed value of $\eta$, removing the regularization by sending $r_w \rightarrow \infty$ in units of $R_+$ puts us eventually in the regime $\eta \,r_w^d \gg1$, i.e. the endpoint is a large AdS black hole of ever growing horizon radius. The black hole is always well-contained inside the sphere of radius $r_w$,
unless we push the parameters to extreme limits, $\eta \sim 1$, in which case the black hole has radius of order $r_w$. In either case, the crunch singularity inside the black hole can be seen as  a regularized version of the boundary crunch singularity that develops in the strict $r_w =\infty$ situation.

What was discussed so far is the expected endpoint of evolution for a system with just two `levels', given by the two AdS$_\pm$
solutions. In explicit CFT realizations, the large $N$ limit actually puts at our disposal $\CO(N)$ metastable vacua obtained by gradually removing constituents branes from the initial AdS$_+$ background. Hence, as progressive nucleation of ever more curved AdS bubbles proceeds, the final curvature of the AdS$_-$ cores becomes eventually of order one, and a geometrical picture breaks down, together with the single-field description based on the $\phi$ collective coordinate. The internal metric for $r<r_w$ having curvature of order one, the threshold at $r_w$ is akin to a `wall', likely to be interpreted as a mass gap and a trivial IR limit of the theory.  Hence, we expect the final endpoint to be a thermal excitation of a massive phase. It would be interesting to elucidate the fate of the global and gauge symmetries, as the branes `pile up' at $r=r_w$, depending on the dynamics on the extra dimensions, such as the ${\bf S}^5$ of the basic AdS/CFT example.  

\bigskip
\centerline{\epsfxsize=0.32\hsize\epsfbox{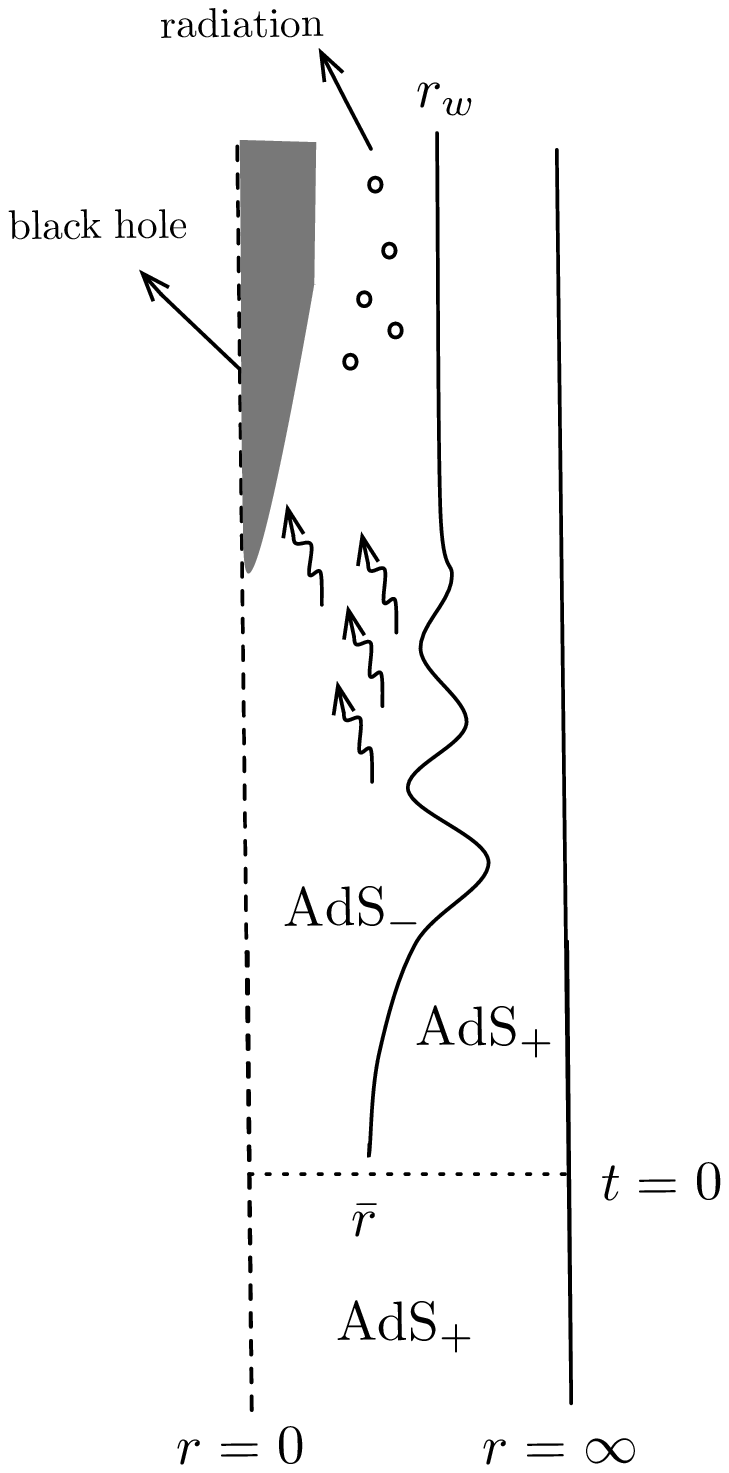}}
\noindent{\ninepoint\sl \baselineskip=8pt {\bf Figure 16:} {\ninerm
Picture in global coordinates showing the nucleation and relaxation of a bubble centered at $r=0$. 
The bubble with ${\rm AdS}_-$ inside  is nucleated at $t=0$ to enter a damped oscillation, relaxing  at $r=r_w$ after the excess energy is radiated away, and finally ending in a radiation equilibrium state which may collapse into a black hole depending on the energetics. The detailed process of the thermalization is actually dominated by a chaotic foam of bubble collisions, rather than this simple picture of single-bubble relaxation.
}}
\bigskip

\subsec{Toy models of potential stabilization}

\noindent

In this section we conclude with some speculations about possible realizations of the stabilization scenario described above. 
Since the key to the bulk description studied in this paper is the existence of a conformally invariant barrier $-\lambda \,\phi^{d/\Delta}$, a natural stabilization would involve adding an irrelevant operator with a high scale $M$ and {\it positive} coefficient (cf. \refs{\eli,\hcc}).  One may imagine engineering the physics at this threshold in such a way that the leading operators below the scale $M$ are a stable irrelevant operator and an unstable marginal operator, i.e. $-\lambda \,\phi^{d/\Delta} + \phi^{d/\Delta + \delta} / M^\delta$, with $\delta >0$. This potential produces a stabilization scale at $\phi_w \sim M \,\lambda^{1/\delta}$. 

This has the difficulty of leaving the UV definition of the theory undetermined, and thus removes the justification for the use of AdS$_+$ as an initial state. Therefore, such a strategy (see e.g. the last example in \eli) must be taken to completion by specifying the detailed physics at the scale $M$ in such a way that the theory flows to an UV fixed point above $M$ and indeed generates the required pattern of signs for the effective operators below $M$. If these conditions are met, we can associate
the UV fixed point above $M$ with the AdS$_+$ state. The detailed geometrical implementation of the stabilization would involve a moderate hierarchy between the minimum of the potential $r_w$ and the transition to AdS$_+$ geometry at $r\sim r_M > r_w$. This makes the model more complicated, but on the other hand all of the qualitative features discussed 
in this section would be realized.

A more self-contained strategy of stabilization can be envisaged by adjusting CFT perturbations using finite-$N$ effects. Let us concentrate on the `template' model of ${\cal N}=4$ SYM in four dimensions with gauge group $SU(N)$. We have seen in
section 4.1 that the operator $\phi^4$ can be generated with a coefficient of $O(1/N)$ from the canonically normalized gauge invariant operator ${1\over N} \Tr\,\Phi^4$. The same operator $\phi^4$ may be generated by projecting the double-trace operator ${1\over N^2} \left(\Tr \Phi^2\right)^2$, now with a coefficient of $\CO(1/N^2)$  (we again assume canonical normalization of the $\Phi$ matrix field.) 
Let us then consider a perturbation of the SYM Lagrangian specified by
\eqn\pertcft{
{\cal L} = {\cal L}_{\rm SYM} - {g_1 (\mu) \over N} \left(\Tr \Phi^4 + \dots\right) + {g_2 (\mu) \over N^2} \left(\left(\Tr \Phi^2 \right)^2 + \dots \right)\;,
}
where $g_1 (\mu)$ and $g_2 (\mu)$ are both  positive at a conventional renormalization scale $\mu$.  The dots stand for extra terms, such as contributions from fermion fields or the detailed $SO(6)$ structure. In particular, we can choose the single-trace operator to be a supersymmetric F-term deformation with generic coupling, which lifts the Coulomb phase of the SYM theory (cf. \refs\ofer) and thus serves as a good stabilization operator in the UV. The double-trace operator breaks supersymmetry and the running of $g_2 (\mu)$ breaks conformal symmetry to one-loop order. The effective coupling being unstable, it is actually asymptotically free, i.e. we have $g_2 (\mu) = c/b_0 \log(\mu/\Lambda_{\rm IR})$, where $b_0 >0$ is the leading beta-function coefficient of the coupling $g_2 / c$, and $\Lambda_{\rm IR}$ the strong infrared scale.   We may further assume $\Lambda_{\rm IR} < 1$, in units of $R_+$, in order for the leading logarithmic running to be a good approximation down to the mass gap scale $1/R_+ =1$.

On further evaluating \pertcft\ along the $U(1)$ direction $\Phi = {\rm diag}\,(0, 0, \dots, \phi)$,  using $\mu=\phi$ as the renormalization scale, we obtain the effective potential 
\eqn\effpphi{
V(\phi) = {g_1 \over N} \,\phi^4 -  {c\,\phi^4 \over N^2\,b_0\,\log(\phi/\Lambda_{\rm IR})}\;,
}
with $c=2\pi^2 /3$ and $b_0 =3$ at one-loop order. This potential  has the required qualitative features with a stabilization scale of order
\eqn\estesc{
\phi_w \sim \Lambda_{\rm IR} \,\exp\left({c \over b_0 \, N\,g_1}\right)\;.}
This expression is corrected at next to leading order both from the renormalization of $g_2$ and the running of $g_1 (\mu)$, which starts once the mixing with the double-trace operator is taken into account. The asymptotic freedom of $g_2$ suggests that these running effects on $g_1$ are negligible at large $\phi$. 

We can now make contact with the notations of section 4. The stabilizing F-term has a coupling with the natural order of magnitude in the large $N$ limit, i.e. $g_1 = \CO(1)$ as $N\rightarrow \infty$. On the other hand, the double-trace operator has a coefficient of order $g_2 / N^2$ with $g_2 = \CO(1)$. Since the natural scale of the $\phi^4$ operator is $(1-\alpha)/N$, according to \effcc, we must set $1-\alpha \sim g_2 /N$ in order to match the potentials. Therefore, we conclude that this stabilization method can only work if implemented at large, but {\it finite} values of $N$. 

The conformal symmetry of the unstable section of the potential   \effpphi\ is spoiled by the logarithmic running of the double-trace coupling. The main consequence regarding the theory of Fubini instantons is the expected lifting of the instanton moduli space, namely the measure for small instantons \meass\ gets replaced by 
\eqn\instlambda{
d^4 x_0  \,{d\rho \over  \rho^{5}} \exp\left(-2\pi^2 N^2 /3g_2 (\rho)\right) = d^4 x_0 \, {d\rho \over \rho^{5}} \,\left(\rho \Lambda_{\rm IR} \right)^{b_0 \,N^2} \;.
}
At large, but finite $N$, the small instanton singularity has been tamed, much as in the Yang--Mills theory, and 
replaced by a large-instanton divergence, which in our case is cured by the `small volume'  assumption  $R_+  \Lambda_{\rm IR} <1$. 

The structure outlined here is based on perturbative intuition, but the main ingredients should hold at strong 't Hooft coupling, if the double-trace perturbation is implemented along the lines of \refs\dtrace, which in particular preserves the one-loop form of the running for the double-trace coupling. The analysis of D3-brane probes in \refs\craps\ and \refs\ofer\ is indeed compatible with our picture. 

A given CFT is associated to a particular set of backgrounds with a given asymptotic behavior. On the other hand, solutions of the `supergravity potential' $U(\sigma, \chi)$, with different asymptotic cosmological constant are not in general related by well-defined tunneling transitions.  Hence, it would seem  that the landscape is `agnostic', i.e. it will  by itself  allow many different CFT models to describe different modes of decay which, from the point of view of the effective potential $U(\chi, \sigma)$, look
 roughly on the same footing. So, regularized models in which the decay process occurs inside a fixed Hilbert space do exist, but the same landscape potential {\it also} contains the decay modes which take place inside the Hilbert space of  sick CFT duals.

\newsec{Discussion}

\noindent

The issues tackled in this paper are rather challenging.  They touch upon the resolution of gravitational singularities in string theory and the very nature of the so called `landscape' \refs\sussk.
In retrospect the authors need  to separate what was well known and/or  self-evident from  the new insights that were gained in this work.
This task can be no less demanding.
The topics themselves have been examined in the past, in particular, as we have already mentioned, this is true of  the idea that the big crunch singularity is in some way or another  related to an instability in a boundary field theory.
In this work  we have set up and studied these questions in the AdS/CFT framework, an arena in which one has somewhat more control.
The basic strategy was to obtain the dual holographic formulation, allowing us  to be more precise in formulating the issues involved,  as well as finding ways to resolve various instabilities.
After that was done one needed to return to the less familiar bulk part and uncover where the medicine offered on the boundary leads  to in the original bulk problem.
We recall several of the important results.

In the issue of a big crunch we have used the observation of Coleman and de Luccia that in some region of parameter space the decay of a false vacuum in AdS space leads to a crunch. We have essentially derived the CFT dual of
the background which is a CdL bubble.  The result is a CFT on the boundary which has no ground state in those cases for which a tunneling can occur and is stable otherwise. We have related the parameters of the bulk theory
with those of the boundary one, checking that the presentation of the dual theory is faithful. The most striking of these  checks was obtained by
successfully reinterpreting the basic physics of the AdS decay, as described by the work of Coleman and de Luccia, in terms of  analogous instanton processes in unstable CFTs. The duality involves a  precise correspondence  of the leading decay rate on both sides, including the detailed identity of the CdL bounce action and the QFT instanton action. This was shown in the limit in which the unstable potential is weakly coupled.
This precise matching is quite intriguing  on its own, since the required semiclassical objects only exist in situations where supersymmetry is broken; perhaps  conformal symmetry is the `\'eminence grise' behind this phenomenon.  

We then discussed the main implications of this result first for  the interpretation and then for the possible `resolution' of the ensuing crunch singularity.
When faced with a crunch singularity one may expect the nonperturbative stringy resolution to either smooth it out, or rather reveal it as an ill-posed question. In the second case one can still  blame the initial state as being unphysical, or  the exact Hamiltonian for being ill defined.

In our case we find that metastable  AdS backgrounds  can indeed  admit  dual descriptions as time-dependent solutions of an unstable Hamiltonian with  no ground state. However, once the CFT was found one could  also say that the initial state was unphysical, having infinite energy with respect to the `true' ground state.  This infinite energy does not reveal itself as the ADM mass of the original AdS$_+$ spacetime but rather, to make sense of the energetics we must  regularize the system so that the dual QFT does have a stable ground state, controlled by a threshold energy scale $\phi_w$, or the corresponding radial scale $r_w$ in the bulk.  When this is done we find that no finite-energy decay process can change the asymptotics of the initial AdS$_+$ spacetime, which stays protected for $r>r_w$. 
This result addresses the concerns raised by T. Banks on the interpretation of these decays \refs\banks.

We then conclude by drawing in bold strokes the endpoint of the decay: a locally thermalized state that results from the `reheating' of the oscillating bubbles by dissipation and collision. In bulk terms, this is a black hole which stays well-contained in the region $r< r_w$ and harbors all the reheating energy. 
Hence, we see that the crunch, if it exists, is safely cloaked behind a black hole horizon. In the limit where we remove the regularization, $r_w \rightarrow \infty$, we obtain the previous cosmological crunch as the black hole grows to infinite size, engulfing the whole background manifold validating yet again the picture we suggested for the unregularized decay.

We have went further and attempted to draw a five dimensional caricature of the resolved decay. This involved a domain wall in which the parameters allowed decay in one region of space but not in the other. For this we needed to allow the participation of several fields in the low energy effective theory.

This leads to some final comments on the landscape. We have started by recalling that in the presence of gravity when one exhibits an effective potential what one sees is not always
what one gets. What would look as local maxima or meta stable minima, had the potential described a field in a QFT with the presence of gravity,  are in fact stable.
Moreover when one considers the potential with four minima we recalled above we have found that the landscape is what we called `agnostic'. The same low energy potential allows both well defined and ill defined CFT's to represent various decays, the effective action is not a very useful diagnostic of which is which, one needs a more extensive analysis to be able to do that.
However the bottom line is that some are well defined and show embryo big crunch features, as much as the theory will allow.

\bigskip{\bf Acknowledgements:}  We are indebted to O. Aharony,  T. Banks, R. Emparan, D. Harlow, J. Mart\'{\i}nez-Mag\'an, N. Seiberg, S. Shenker, E. Witten for useful discussions. We would like to thank the Perimeter Institute for their kind hospitality.  The work of J.L.F. Barb\'on was partially supported by MEC and FEDER under grants FPA2006-05485 and FPA2009-07908, the Spanish
Consolider-Ingenio 2010 Programme CPAN (CSD2007-00042) and  Comunidad Aut\'onoma de Madrid under grant HEPHACOS P-ESP-00346. The work of E. Rabinovici was partially supported by the Humbodlt foundation, a  DIP grant H, 52, the Einstein Center at the Hebrew University, the American-Israeli Bi-National Science Foundation and the Israel Science Foundation Center of Excellence.

\bigskip{\bf Note Added:} After this paper was posted, we became aware of previous work \refs{\tassos, \tpapa, \papa} containing   significant overlap with  section 4.

\newsec{Appendix: Thin Euclidean bounces with $O(d+1)$ symmetry}

\noindent

The Hamiltonian formalism used in this paper is adapted to the boundary geometry ${\bf R} \times {\bf S}^{d-1}$ and it is rather transparent from the physical point of view, illustrating the different
phenomena at play, such as barrier penetration, Schwinger production, or Hawking emission by a black hole. The associated Euclidean bounces have $U(1)\times O(d)$ symmetry. However, when discussing the purely vacuum transitions with $M_+ = \omega=0$, such as the basic CdL process, it is useful to exploit the full symmetry of the problem and work with a formalism with manifest $O(d+1)$ symmetry, which in addition is helpful in making contact with the classic results of \refs\cdl. 

A form of the vacuum AdS$_+$ metric with manifest $O(d+1)$ symmetry is the Poincar\'e ball:
 \eqn\odmll{
 ds^2_+ = d\rho^2 + \sinh^2 (\rho)\,d\Omega_d^2 = {dr^2 \over 1+r^2} + r^2 \,d\Omega_d^2\;,
 }
 and the $O(d+1)$ symmetric bounce in the thin-wall approximation is simply given by an embedded ${\bf S}^d$
 sphere with fixed radius $r$. The corresponding Euclidean action is 
\eqn\eucss{
I_E = \sigma\,{\rm Vol} \left[\,{\bf S}^d\,\right] - (qd) \,{\rm Vol}_+ \left[\,{\bf B}^{d+1}\,\right]\;,
}
where ${\bf B}^{d+1}$ stands for the $(d+1)$-ball conforming the  {\it interior} of ${\bf S}^d$ with its volume  computed in the {\it exterior} AdS$_+$ metric. The nucleation rate is proportional to   $\exp(-I_E)$, with the action \eucss\ evaluated at a local maximum in the space of brane configurations.

 Evaluating the Euclidean action for a spherical Euclidean brane of radius $r= \sinh(\rho)$ we find
 \eqn\evale{
 I_E (r) = \sigma \,|{\bf S}^d | \,r^{d} - d\,q\,|{\bf S}^d |\, \int_0^r {dz \,z^d \over \sqrt{1+z^2}}\;,
 }
 which indeed has a local maximum at $r={\bar r}= \alpha/\sqrt{1-\alpha^2}$. The extremal value of the action is given by 
 \eqn\accee{
 I_E = q\,|{\bf S}^d | \,{\bar r}^{\,d+1} \,\int_0^1 {dy \,y^d \over (1+{\bar r}^2 y^2)^{3/2}}\;,
 }
where we have used $\sigma = \alpha \,q$ and applied an integration by parts in the second term in \evale. On further evaluating the definite integral in terms of hypergeometric functions one finds 
\eqn\hpp{
{(d+2) F\left[\sdmasonehalfs, -\shalf, \sdmasthreehalfs, -{\bar r}^{2}\right] - (1+d({\bar r}^2 +1)) F\left[\sdmasonehalfs, \shalf, \sdmasthreehalfs, -{\bar r}^{2}\right] \over (d+1)( 1+ {\bar r}^{2})}\;.
}
This expression  can be further reduced into
\eqn\reddu{
I_E = q\,|{\bf S}^d|{{\bar r}^{\,d+1} \over d+1} {1\over \sqrt{1+{\bar r}^{\,2}}} \,F\left[1, \sdhalfs, \sdmasthreehalfs, -{\bar r}^{\,2}\right]\;,
}
which is equal to the Hamiltonian form of the integral in \we\ (recall $v=|{\bf S}^{d-1}|$)  
\eqn\hamii{
I_E = 2W_E = 2\,q\,v\,{{\bar r}^{\,d+1} \over \sqrt{1+{\bar r}^{\,2}}} \int_0^1 {dx\,x^{d-1} \over 1+ {\bar r}^{\,2} x^2} \sqrt{1-x^2}\;.
}

The very explicit forms of the bounce action allow us to check some of the results of the original CdL work \refs\cdl. In particular, we can look at a large-bubble regime which was not discussed so far in this paper, namely the situation where $R_- \ll {\bar r} \ll R_+$. This corresponds to the decay of a flat Minkowski space by nucleation of large AdS bubbles, hence the interest of \refs\cdl\ on this particular case. Restoring the dependence on $R_+$ in the general expression \hamii\ and taking the $R_+ \rightarrow \infty$ limit we find
\eqn\lbb{
I_E \rightarrow {q |{\bf S}^d | \over d+1}\; {\bar r}^{\,d+1}\;.
}
In order to make contact with the notation in \refs\cdl\ we parametrize the bounce action in terms of the corresponding quantities in the absence of gravitation, i.e. the nucleation size:
\eqn\uus{
{\bar r}_0 = {\sigma \over q_0} = {\sigma d \over \Delta U} = {2\kappa \over \Delta \lambda} \;,
}
and the bounce action
\eqn\uub{
B_0 = {|{\bf S}^d | \over d+1} \,q_0\, {\bar r}_0^{\,d+1}\;.
}
In the limit $R_+ \rightarrow \infty$ we have ${\bar r} = \sigma /q$ and
$$
q=q_0 \left(1-\left({{\bar r}_0 \over 2 R_-}\right)^2 \right)\;.
$$
Hence we can rewrite \lbb\ as
\eqn\lldd{
I_E = {B_0 \over \left(1-({\bar r}_0 /2R_-)^2 \right)^d}\;,
}
 which disagrees with the $d=3$ result of \refs\cdl\ by a factor of $1-({\bar r}_0 / 2R_- )^2$. The general expression for the $O(4)$-invariant bounce action in \refs\cdl\ takes the form (adapted to our notations)
 \eqn\cdlf{
 B({\bar r}) = \sigma\,|{\bf S}^3 |\, {\bar r}^{\,3} - {|{\bf S}^3 | \over 4\pi G} \left[R_-^2 \left(\left(1+{{\bar r}^2   \over R_-^2} \right)^{3/2} -1\right) - R_+^2 \left( \left(1+{{\bar r}^2 \over R_+^2} \right)^{3/2} -1 \right)\right]\;
 }
The nucleation size ${\bar r}$ corresponds to the local maximum of this function, i.e. to the solution of the equation 
\eqn\loccma{
0=B' ({\bar r}) = {3{\bar r} \over 4\pi G} \left(\kappa\,{\bar r}  +\sqrt{f_+ ({\bar r})} - \sqrt{f_- ({\bar r})} \right)\;,
}
which is indeed satisfied at the turning point $dr/d\tau =0$, according to \roots. Despite providing the right value of ${\bar r}$, the 
expression \cdlf\  differs in general  from our  result \accee. The ratio $I_E ({\bar r}) / B({\bar r})$, computed as a function of ${\bar r}$ and at fixed $R_\pm$, approaches unity for small bubbles, ${\bar r} \ll R_\pm$, but increases monotonically until the 
 mismatch stabilizes at $R_+ /R_-$ for very large bubbles, i.e. we find 
$$
B({\bar r}\gg 1) \longrightarrow {|{\bf S}^3 |  \over 4\pi G} \,(R_+ - R_-)\,{\bar r} \longrightarrow {R_- \over R_+} \,I_E\;.
$$
Hence, the matching to Fubini instantons is compromised by a factor of $R_- /R_+$, should we use the expression of  \cdl. For the case of ${\cal N}=4$ SYM, this mismatch is admittedly small in the large $N$ limit, since the effective
coupling $\lambda$ is itself of order $1/N$ (in this case $R_- /R_+ \approx 1-1/N$).  However, it must be said that the detailed matching presented in section 4 only required $\sigma \rightarrow q$, with no constraints placed on the value of $R_+ / R_-$. 
At any rate, these  considerations show that the detailed treatment of the thin-wall approximation in \refs\cdl\ is not completely equivalent to the one following from the junction conditions, nor is it capable of a perfectly successful  AdS/CFT matching.

\listrefs

\bye